\documentclass[aps,twocolumn,superscriptaddress,nobibnotes,showkeys, english]{revtex4-2}
\usepackage{times}
\usepackage{booktabs}
\usepackage[utf8]{inputenc}
\usepackage{amsmath}
\usepackage{amssymb}
\usepackage{graphicx}
\usepackage{bm}
\usepackage{subfigure}
\usepackage{color}
\usepackage[normalem]{ulem} 
\usepackage{dcolumn}
\usepackage{xr}
\usepackage{hyperref}
\usepackage{cleveref}
\usepackage{float}
\usepackage{orcidlink}
\renewcommand{\vec}[1]{\mathbf{#1}}
\renewcommand{\Re}{\operatorname{Re}}
\renewcommand{\Im}{\operatorname{Im}}
\newcommand{\sgn}[1]{\operatorname{sgn}\left(#1\right)}

\newcolumntype{d}[1]{D{.}{.}{#1}}
\newcommand{\arccosh}{\operatorname{arccosh}}

\newcommand{\Ccont}{\mathcal{C}}

\newcommand{\subdiel}{_{\rm d}}

\newcommand{\subshell}{_{\rm s}}

\newcommand{\supsct}{^{\rm sct}}
\newcommand{\subsct}{_{\rm sct}}

\newcommand{\colvec}[2]{ \left( \begin{matrix} #1 \\ #2 \end{matrix} \right)}
\newcommand{\mat}[4]{ \left( \begin{matrix}  #1 & #2 \\ #3 & #4 \end{matrix} \right)}

\begin{document}


\title{Nonconcentric Multi-shell Nanowires: Geometry-Induced Plasmon Hybridization and Near-Field Localization}




\author{Gino Wegner\,\orcidlink{0000-0001-6225-5269}}
\affiliation{Humboldt-Universität zu Berlin, Institut für Physik, 
	AG Theoretische Optik \& Photonik, 12489 Berlin, Germany}
\affiliation{Institute of Condensed Matter Theory and Optics, 
	Friedrich-Schiller-University Jena, Max-Wien-Platz 1,
	07743 Jena, Germany}
\affiliation{Max-Born-Institut, 12489 Berlin, Germany}
\author{Jer-Shing Huang\,\orcidlink{0000-0002-7027-3042}}
\affiliation{Leibniz Institute of Photonic Technology, 
	Albert-Einstein Str. 9, 07745 Jena, Germany}
\author{Kurt Busch\,\orcidlink{0000-0003-0076-8522}}
\affiliation{Humboldt-Universität zu Berlin, Institut für Physik, 
	AG Theoretische Optik \& Photonik, 12489 Berlin, Germany}
\affiliation{Max-Born-Institut, 12489 Berlin, Germany}

\date{\today}

\begin{abstract}
Localized surface plasmons (LSPs) in multi-shell nanostructures provide a versatile route for controlling optical fields at the nanoscale, yet the influence of deviations from concentric geometries remains insufficiently understood. Here, we investigate the impact of shell nonconcentricity on the quasistatic optical response of core-single-shell and core-multi-shell nanowires. Exploiting the conformal properties of bipolar coordinates, we derive analytical solutions for nonconcentric cylindrical interfaces and systematically analyze the evolution of LSP resonances, absorption spectra, and near-field distributions. Starting from single-shell structures, we show that nonconcentricity enables finite coupling of incident radiation to higher-order plasmon modes that are optically inactive in the quasistatic concentric limit. Extending the analysis to multi-shell bull's eye wires, we identify how shell thickness, number of fixed-thickness shell units, each defining a set of a dielectric and metal shell, as well as interface nonconcentricness shape the hybridized plasmon spectrum. Increasing the number of metal-dielectric interfaces broadens the spectral response, while nonconcentric geometries additionally increase the density of accessible resonances and localize electromagnetic fields preferentially within and around the thinner shell sections. Eventually, comparison of concentric with bipolar and Doppler-grating-inspired nonconcentric bull's eye wires based on Mie theory and full-wave Discontinuous Galerkin Time-Domain simulations, respectively, allows to assess the impact of nonconcentricness for typical nanowire dimensions. These results provide insight into geometry-induced plasmon hybridization and suggest routes toward nanoscale control of optical energy localization for applications in active nanophotonics and plasmon-assisted photochemistry.
\end{abstract}


\maketitle

\section{\label{sec:introduction}Introduction}

Over the past decades, the study of localized surface plasmons (LSPs) has enabled a wide range of applications, including gas sensing~\cite{Proenca2024,Chen2025}, single-molecule detection~\cite{Anker2008,Kanehira2023,Yu2020singlemolSERSroadmap}, fluorescence enhancement and quenching~\cite{PhysRevLett.96.113002}, and nanoscale topographic manipulation~\cite{Na2008}, to name only a few. Besides the choice of the constituent materials, both the overall dimensions of a plasmonic structure and, in particular, the geometry of its conductor--dielectric interfaces govern its optical response. Since the collective charge oscillations underlying LSPs are confined to these interfaces, even subtle geometrical modifications can induce pronounced changes in resonance frequencies and near-field distributions. This sensitivity originates from the self-consistent interaction of the oscillating surface charges~\cite{Inglesfield1973,Ehrenreich1959,Blinder1965}.

Although modern nanofabrication techniques permit highly accurate control over nanostructure geometries, deviations from the intended design can arise during fabrication~\cite{cui2010,Luo2013,Zhang2010}. One such geometry is the nonconcentric alignment of shells surrounding a nanowire core, which forms the central subject of the present work. Importantly, such nonconcentric configurations need not result solely from fabrication imperfections; the precision offered by state-of-the-art nanofabrication techniques also enables their intentional realization as designed nanostructures. In contrast to stochastic imperfections such as surface roughness~\cite{Loth_23,Oldenburg1998,Zhao2024}, nonconcentric shell geometries can be described analytically under suitable assumptions, while providing a model for both controlled geometric configurations and a relevant class of fabrication-induced deviations. In the present work, we investigate the influence of such nonconcentricity on the quasistatic LSP response of both single- and multi-shell nanowires. Specifically,
we consider infinitely long nanowires with cross-sectional dimensions ranging from a few to several tens of nanometers and solve the governing quasistatic boundary-value problem in bipolar coordinates~\cite{morse1953methods,LuchtBipolar}. Bipolar coordinates are conformally related to Cartesian coordinates, allowing the Laplace equation to retain its form while naturally describing nonconcentric cylindrical interfaces. More generally, conformal transformations constitute an important tool in transformation optics, where they provide physical insight into electromagnetic fields at deeply subwavelength scales. In particular, they reveal how different plasmonic geometries may share common spectral properties and establish connections between the continuity or discreteness of plasmonic spectra and the presence or absence of geometrical singularities~\cite{Pendry2012_TO_review}.
Transformation-optics-inspired plasmonic structures frequently exhibit exceptional light-concentrating capabilities by compressing optical energy from wavelength-scale illumination into nanoscale volumes. Such structures therefore represent efficient light-harvesting devices whose performance is often optimized through careful tuning of their geometrical parameters~\cite{Pendry2012_TO_review,Aubry_2010_NanoLett,Lei_2010,Aubry_2010_PRB,PhysRevLett.105.233901,surf_plas_and_singularities,Luo2013}. Motivated by these developments, we investigate how nonconcentric shell geometries modify the spectral and spatial characteristics of localized surface plasmons in analytically tractable nanowire systems.

To elucidate the role of the individual geometrical parameters, we construct the final multi-shell nanowire from a sequence of progressively more complex building blocks. Since each additional shell introduces new geometrical degrees of freedom, this approach allows the influence of each parameter on the quasistatic LSP spectrum and the associated field distributions to be assessed systematically.
We begin with core--single-shell nanowires consisting of either a metallic core coated by a dielectric shell or a dielectric core enclosed by a metallic shell. These configurations isolate the influence of the inner and outer shell radii on the localized plasmon resonances and provide the foundation for understanding more complex structures.
Subsequently, we investigate multi-shell nanowires comprising a metallic core surrounded by a finite number of shell units. Each unit consists of a dielectric, followed by a metallic shell. The total radial thickness of the unit is fixed and the relative thicknesses of the dielectric and metallic layers can be fixed to a value different from unity. The chosen thickness ratio and total thickness will be the same for each unit. The simplest realization, containing a single shell unit, corresponds to the \emph{tube-wire cavity}~\cite{Luo2013,YuLuo_2012}. Owing to the coexistence of plasmons supported by the metallic core and the outer metallic shell, this geometry provides a natural platform for interpreting the optical response within the plasmon-hybridization framework~\cite{Prodan_2003_hybrid_first,Nordlander2004}. Increasing the number of shell units successively introduces additional conductor--dielectric interfaces and consequently additional hybridized plasmon modes. These multi-shell structures, which we refer to as bull's eye wires, therefore provide a versatile platform for studying increasingly rich mode interactions and spatially separated plasmonic hot spots.
Besides concentric structures, we investigate two distinct realizations of nonconcentric multi-shell nanowires. In the first, referred to as the \emph{bipolar bull's eye wire}, all interfaces coincide with coordinate lines of the bipolar coordinate system. In the second, denoted the \emph{Doppler bull's eye wire}, the shell boundaries instead follow the wavefronts emitted by a moving point source and therefore correspond to the cross section of a Doppler grating~\cite{See2017}. Comparing these two geometries enables us to distinguish the influence of different classes of nonconcentric shell arrangements on localized surface plasmons.

The demonstrated differences between both nonconcentric wire assemblies motivate a versatiliy of Doppler wires (and possibly gratings) lying beyond contemporary achievements based on propagating surface plasmon polaritons (SPPs) involving a hybrid-LSP-related hotspot localization within selected shells. Such selective field confinement could provide new opportunities for plasmon-assisted photopolymerization restricted to chosen shells or for the targeted excitation of optically active constituents, such as fluorescent dyes, embedded within individual dielectric layers. In this sense, the present work explores the potential of Doppler-grating-inspired geometries to serve not only as platforms for manipulating propagating surface waves but also as versatile nanoscale near-field engineering devices.
Given the increasing size of the bull's eye wires, we finally conduct full-wave calculations for the concentric structures based on the analytical Mie theory and for the nonconcentric structures based on the discontinuous Galerkin time-domain (DGTD) finite-element method~\cite{lpor_DGTD_review}.

Beyond their fundamental interest, the structures considered here are relevant for plasmon-assisted photochemistry and active nanophotonic systems. In particular, dielectric coatings incorporating optically active constituents, such as fluorescent dyes, have been realized experimentally on gold nanospheres~\cite{Khitous2023,Khitous2023PlasmonInducedPO} and nanorods~\cite{Stete2017}. Related core--shell nanoparticles without embedded emitters have likewise been fabricated in both concentric~\cite{Oldenburg1998,Brongersma2003} and nonconcentric~\cite{Lu2005} configurations, while metallic nanoparticles comprising multiple concentric shells have also been demonstrated experimentally~\cite{Prodan_2003_hybrid_first}. 
We expect that shell-interface nonconcentricity may significantly modify the spatial distribution of plasmonic near-fields and thereby influence the coupling of emitters to localized surface plasmons.
Similar considerations apply to near-field-induced photopolymerization, where distortions of the optical near field may affect the growth of additional shells on an initially nonconcentric precursor. In this context, we note the combined experimental and numerical study of Ref.~\cite{Khitous2024}, which investigated the influence of metallic surface geometry on photopolymerization around nanodisks of circular, hexagonal, and triangular cross section as well as around nanorods.

Throughout this work, we keep in mind the limitations of the quasistatic approximation. In particular, it is known to overestimate near-field enhancements and absorption efficiencies at plasmonic resonances~\cite{Aubry_2010_NanoLett,Lei_2010,Aubry_2010_PRB,PhysRevLett.105.233901,PhysRevB.82.205109}. Nevertheless, the quasistatic approach provides leading-order intuition regarding the dependence of localized surface plasmon resonances and modal field distributions on geometry, thereby facilitating mode identification in full-wave Maxwell simulations (see, e.g., Ref.~\cite{Moeferdt2018}). Accordingly, our discussion primarily focuses on resonance frequencies and spatial field distributions rather than on quantitative values of field enhancement or absorption efficiency.

The remainder of this manuscript is organized as follows. In \cref{sec:governing_equations}, we formulate the quasistatic boundary-value problem and summarize its analytical solution. Section~\cref{sec:core_shell} investigates concentric and nonconcentric core--single-shell nanowires, emphasizing their localized surface plasmon dispersion relations, near-field distributions, and absorption characteristics. The corresponding analysis for multi-shell structures is presented in \cref{sec:multi_ring_systems}. We first examine the tube-wire cavity in \cref{subsec:tube_wire_cavity}, before investigating bull's eye wires containing multiple shell units in \cref{subsec:multi_ring_mu_greater_one}. Finally, \cref{subsec:multi_ring_compare_doppler} compares bipolar and Doppler bull's eye wires and discusses the implications of their distinct nonconcentric geometries for localized surface plasmon resonances and near-field localization.

\section{\label{sec:governing_equations}Governing equations and coordinate systems}

Within the quasi-static approximation, retardation effects are neglected
and the electric field is represented via the gradient of a potential 
function $V$ as
\begin{align}
\vec{E} = - \nabla V 
\label{eq:static_faraday_law}.
\end{align}
For a monochromatic electric field with frequency $\omega > 0$ and a piece-wise
constant material distribution, the quasistatic approximation then leads to
\cite{Ford1984}
\begin{align}
0 
= 
\nabla \cdot \left[ \nabla \times \vec{H}(\vec{r}, \omega)\right]
= 
-i \omega \epsilon(\omega) 
\nabla \cdot \left( \nabla V(\vec{r}, \omega) \right).
\label{eq:div_of_qustatic_ampere_law}
\end{align}
Here, we have introduced linear, nonmagnetic, spatially local and dispersive
material distribution via the dielectric function $\epsilon(\omega)$ such 
that the constitutive relations read 
$\vec{D} (\vec{r}) = \epsilon_0 \epsilon (\vec{r},\omega) \vec{E} (\vec{r}) $ 
and
$\vec{B} (\vec{r}) = \mu_0\vec{H} (\vec{r})$. 

By way of the Maxwell-Amp\`ere law, the dielectric function $\epsilon (\omega)$ 
contains additive contributions stemming from the response of free and bound electrons \cite{Pfeiffer_2024}. 
For definiteness and since our focus lies on the complex geometry, we employ 
for the metallic constituents a dielectric function of Drude type 
\begin{align}
\epsilon(\omega) &= 1 - \frac{\omega^2_{\rm p}}{\omega(\omega+i\gamma)},
\label{eq:Drude}
\end{align}
with plasma frequency $\omega_{\rm p}$ and damping constant $\gamma$. These 
parameters are  obtained from fitting to experimental data. For instance,
data for silver can be found in Ref.~\cite{Johnson1972} and, for this case,
the fit yields $\omega_{\rm p}=9.149$eV and $\gamma=0.021$eV. However, we
would like to note that the direct incorporation of measured data or other 
local material models is straightforward. 

The strategy for obtaining the solution for a composite system described
by a spatially piece-wise constant dielectric functions is as follows.
In every material domain the Laplace equation is solved and subsequently
these local solutions are combined to a global solution via the boundary 
conditions between the different domains \cite{Ford1984}
\begin{align}
	(\hat{t}\cdot \nabla) V|_+ - (\hat{t}\cdot \nabla) V|_-  
	& = 
	0,
	\label{eq:continuous_tang_electric_field} \\
	(\epsilon(\omega)(\hat{n}\cdot \nabla) V)|_+ 
	- 
	(\epsilon(\omega)(\hat{n}\cdot \nabla) V)|_- 
	& = 
	0.
	\label{eq:continuous_norm_diel_displ}
\end{align}
Here, $\hat{t}$ and $\hat{n}$ denote, respectively, unit vectors in
tangental and normal direction at the interface between adjacent 
materials and the superscript $\pm$ refers to the different sides 
of the interface.

\begin{figure}[h]
	\centering
	\includegraphics[scale=.5]{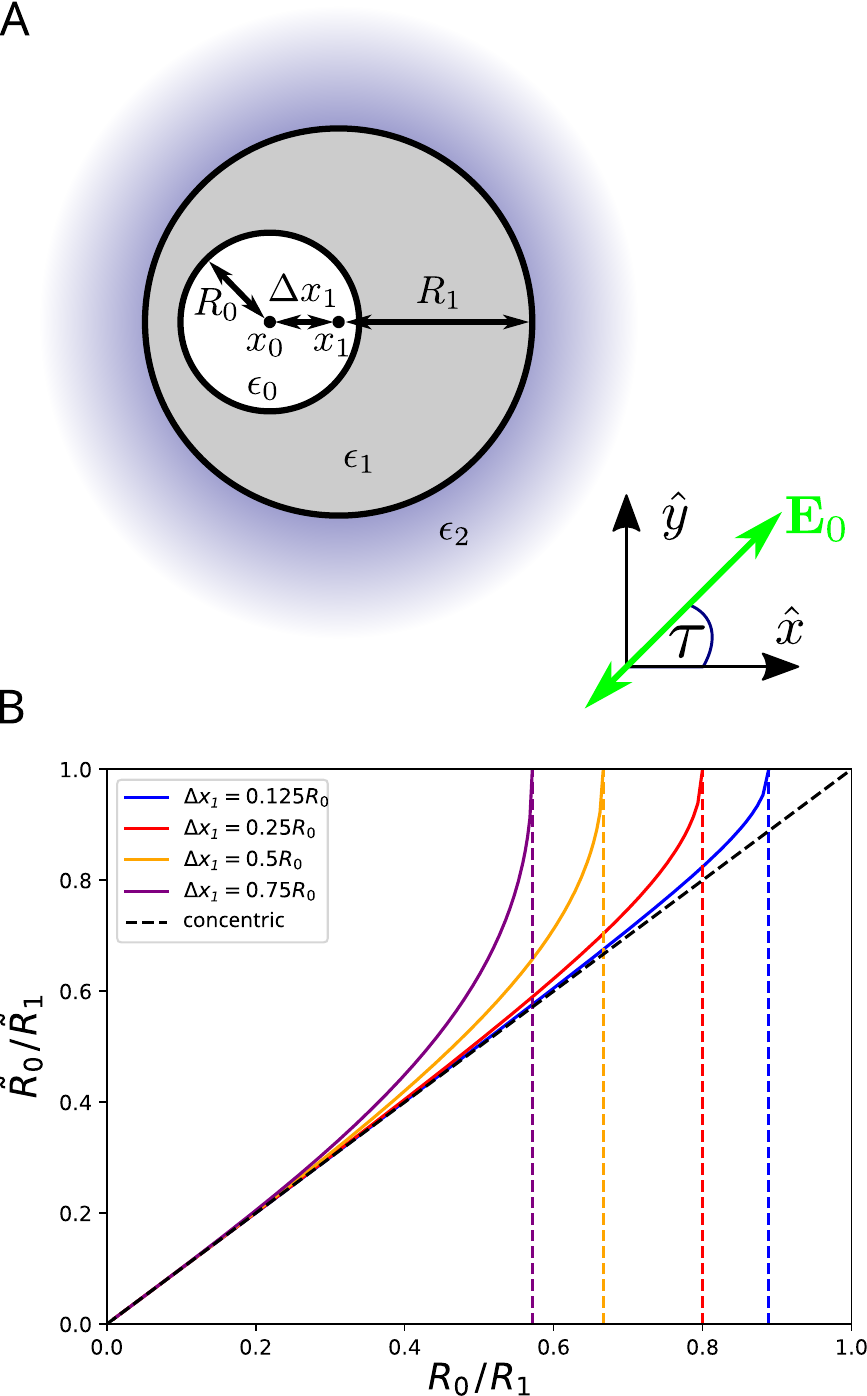}
	\caption{
		\label{fig:core_shell_setup_eff_radius}
		Panel A: Geometry of a core--single-shell nanowire with inner and outer radii $R_0$ and $R_1$, respectively. The shell's outer boundary is displaced by $\Delta x_1=x_1-x_0$, and the three regions are characterized by dielectric functions $\epsilon_\alpha$ ($\alpha\in\{0,1,2\}$). The externally applied electric field $\mathbf{E_0}$ is linearly polarized at an angle $\tau$ with respect to the $x$ axis (green arrow).
		Panel B: Effective radius ratio $\tilde{R}_0/\tilde{R}_1$ as a function of the physical radius ratio $R_0/R_1$ for different displacements $\Delta x_1$. The vertical dashed lines indicate the intersection (crescent) limit, $R_1=R_0+\Delta x_1$, while the black dashed line corresponds to the concentric geometry ($\Delta x_1=0$).
		}
\end{figure}

In the present study, we focus on long metallic wires (oriented
along the $z$-axis) with complex cross-sectional structures (for 
a single shell see \cref{fig:core_shell_setup_eff_radius})
which are illuminated by linearly polarized monochromatic plane waves 
propagating perpendicular to the wire axis.
We incorporate the illuminating plane wave with wave vector 
$\vec{k} = k_x \hat{x} + k_y \hat{y}$
($\hat{x}$ and $\hat{y}$ denote the unit vectors along the axes
tranverse to the wire)
via a far-field condition and, in cartesion coordinates, we
explicitly have
\begin{align}
	\vec{E_0}(\vec{r}, \omega) 
	&= 
	E_0 \left( \cos(\tau) \, \hat{x} + \sin(\tau) \, \hat{y} \right) 
\end{align}
where $\tau=\angle(\mathbf{E_0},\hat{x})$ is the angle between the 	
polarization of the plane wave and the $x$-axis.

For an analytical treatment of complex geometries, it is advantageous 
when the material interfaces follow coordinate isolines so that the 
boundary conditions
(\cref{eq:continuous_tang_electric_field} 
and 
\cref{eq:continuous_norm_diel_displ})
can be solved in a straightforward manner. Given that in this work
we are considering the cross-sections of concentric 
and non-concentric
shells, this suggests that we utilize, respectively, polar
(or equivalently log-polar) and bipolar coordinate systems. 
In \cref{app-sec:potential_ring_structures}, we therefore 
describe the log-polar and the bipolar coordinate systems 
as well as the solutions of the Laplace equation in these 
coordinate systems.

In the remainder of the manuscript the index $\alpha$ enumerates the 
media contained in different shells starting from the core ($\alpha=0$) 
in the center. While the core is bounded by the single radius $R_0$, each 
shell, labeled as medium $\alpha$ with $\alpha>1$ is bounded by a large 
and small circle with radii $R_\alpha$ and $R_{\alpha-1}$. In addition,
these circles may be horizontally displaced by an amount 
$\Delta x_\alpha = x_{\alpha}-x_{\alpha-1}$, where $x_\alpha$ denotes  
the $x$-coordinate of the center of the circle labeled by $\alpha$,
see e.g. \cref{fig:core_shell_setup_eff_radius}~A for the simplest
configuration of a core and a single shell. 
For further details, we refer to \cref{app-sec:potential_ring_structures,sec:Mie_theory}.
For a concentric realization, see, e.g., \cref{fig:Mie_sketch_con_bulls_eye}.

\section{\label{sec:core_shell}Treatment of core-single-shell nanowires}

Before considering more complex multi-shell geometries, we first investigate core--single-shell nanowires consisting of a single shell bounded by two circular interfaces with radii $R_0$ and $R_1$, whose centers are displaced by $\Delta x_1=x_1-x_0$ along the $x$ axis (see \cref{fig:core_shell_setup_eff_radius}~A). While Ref.~\cite{Zhang13} primarily interpreted the optical response of these structures in terms of Fano resonances and proposed them as multiwavelength refractive-index sensors, our focus here is on the geometrical origin of the localized surface plasmon (LSP) spectrum.

The corresponding quasistatic boundary-value problem is solved in \cref{app-sec:potential_core_shell}. The LSP dispersion relations follow from the poles of the scattered-field coefficients (see \cref{app-eq:core_shell_B_e_n} for the metal-core--dielectric-shell configuration and \cref{app-eq:di_core_met_shell_B_e_n} for the dielectric-core--metal-shell configuration). For the metal-core--dielectric-shell case ($\epsilon_0=\epsilon(\omega)$ and $\epsilon_1=\epsilon_{\mathrm{d}}$), the resulting implicit dispersion relation reads
\begin{align}
	\epsilon(\omega_n)
	+
	\epsilon_1\mathcal{Q}_n(\tilde{R}_0/\tilde{R}_1;\tilde{\epsilon}_{12}) 
	= 0,
	\label{eq:impl_disp_rel_met_core_die_shell} 
\end{align}
where $\omega_n$ denotes the resonance frequency of the mode of order $n$ localized at the core interface, $\tilde{\epsilon}_{ij}=(\epsilon_i-\epsilon_j)/(\epsilon_i+\epsilon_j)$, and
\begin{align}
	\mathcal{Q}_n(\tilde{R}_0/\tilde{R}_1;\tilde{\epsilon}_{\rm 12}) 
	& = 
	\frac{1 - (\tilde{R}_0/\tilde{R}_1)^{2n}\tilde{\epsilon}_{\rm 12}}{1 + (\tilde{R}_0/\tilde{R}_1)^{2n}\tilde{\epsilon}_{\rm 12}}.
	\label{eq:f_func_def}
\end{align} 

The effective radial ratio appearing in this expression is given by
\begin{align}
	\frac{\tilde{R}_0}{\tilde{R}_1}
	= \frac{R_0}{R_1}
	\frac{x_1  + \sqrt{x_1^2  - R^2_1 }}
	{x_0 + \sqrt{x_0^2 - R^2_0}}
	\label{eq:ratio_effective_radii}
\end{align}
where $x_0$ and $x_1$ are determined by the radii and displacement. Thus, within the quasistatic description, nonconcentricity affects the LSP spectrum exclusively through the replacement of the physical radial ratio by an effective ratio in the transformed space. This result follows directly from the bipolar-coordinate solution presented in \cref{app-sec:potential_core_shell}. In the concentric limit, $\Delta x_1\rightarrow0$, the effective ratio continuously approaches the physical radius ratio [see \cref{app-eq:eff_rad_a_Dx_to_zero}]. Consequently, the difference between these two ratios provides a convenient, mode-independent measure of the influence of nonconcentricity.

The dependence of the effective ratio on the actual radius ratio is shown in \cref{fig:core_shell_setup_eff_radius}~B for several interface displacements. The largest deviation occurs for $\Delta x_1=R_1-R_0$, corresponding to the intersection limit of the two interfaces [see \cref{eq:crescent_limit}], where the shell adopts a crescent-like shape. Within the Drude approximation, the dispersion relation becomes
\begin{align}
	\omega_n &= \frac{\omega_{\rm p}}{\sqrt{1 + \epsilon_1 \mathcal{Q}_n(\tilde{R}_0/\tilde{R}_1, \tilde{\epsilon}_{12})}}
	\label{eq:noncon_met_core_diel_shell_w_LSP}
\end{align}
provided that $\gamma^2\ll \omega^2_{\rm p}/(1 + \epsilon_1 \mathcal{Q}_n)$.

\begin{figure*}
	\centering
	\includegraphics[scale=0.5]{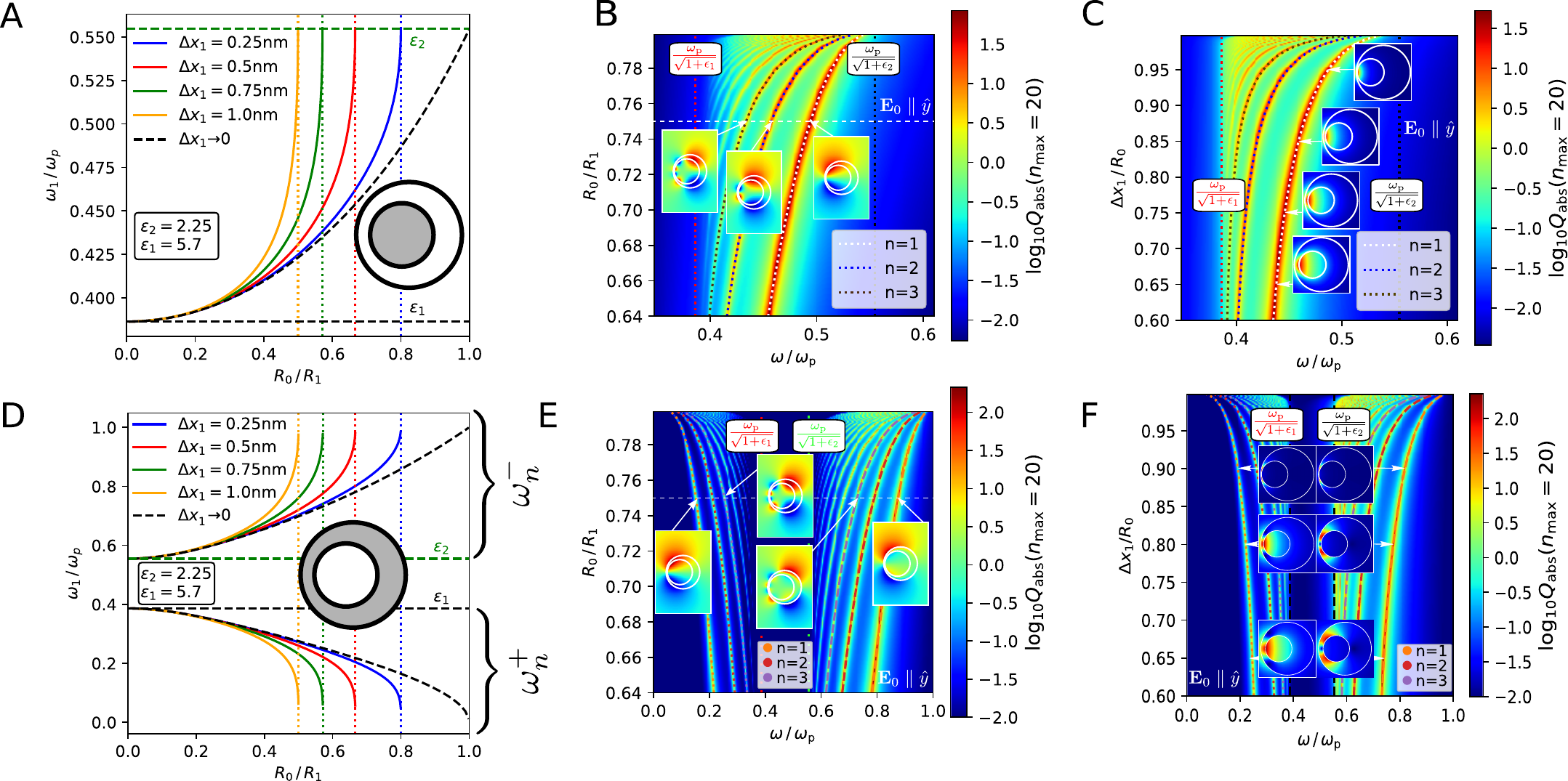}
	\caption{
		\label{fig:core_shell_disp_rel_Q_abs}
		Panels A and D:
		Dipolar LSP resonance frequency as a function of the physical radius ratio for different shell displacements $\Delta x_1$, obtained from 
		\cref{eq:noncon_met_core_diel_shell_w_LSP} 
		and \cref{eq:noncon_diel_core_met_shell_w_LSP}. The  core is surrounded by a diamond shell with dielectric function $\epsilon_1=5.70$\cite{diamond_diel_const} and a silica environment with $\epsilon_2=2.25$ (corresponding to a refractive index of approximately $1.5$).
		Panels B and E: Absorption efficiency as a function of frequency and radius ratio $R_0/R_1$ for a fixed displacement $\Delta x_1=0.25R_0$. The core radius amounts to $R_0=5\,\mathrm{nm}$ and the material parameters correspond to those in panels A and D, with the metallic region described by a Drude model with plasma frequency $\omega_{\rm p}=1.39\times10^{16}\,\mathrm{s}^{-1}$ and damping rate $\gamma=3.23\times10^{13}\,\mathrm{s}^{-1}$. The dotted lines indicate the resonance positions predicted by the corresponding dispersion relations. Insets show the imaginary part of the scattered potential for selected LSP modes given $\mathbf{E_0}\parallel \hat{y}$.
		Panels C and F: Absorption efficiency as a function of frequency and shell displacement for a fixed outer radius $R_1=2R_0$. The core radius and material parameters are identical to those in panels B and E. Insets display the modulus of the scattered electric field at the dipolar resonance for selected displacements given $\mathbf{E_0}\parallel \hat{y}$. 
	}
\end{figure*}

Although this expression contains solutions of all multipolar orders, only the dipolar mode can be excited by a spatially uniform external field in the concentric limit. The remaining solutions correspond to dark quasistatic modes. Nonconcentricity relaxes this symmetry constraint, allowing higher-order resonances to acquire a finite optical response. The dipolar dispersion is plotted in \cref{fig:core_shell_disp_rel_Q_abs}~A as a function of the actual radius ratio for increasing displacement.

For concentric structures, the resonance frequency continuously interpolates between the surface plasmon frequencies associated with the isolated shell interface, $\omega=\omega_{\rm p}/\sqrt{1+\epsilon_1}$, and the isolated environment interface, $\omega=\omega_{\rm p}/\sqrt{1+\epsilon_2}$. Physically, this reflects the finite spatial extent of the plasmonic near field: depending on the geometry, the induced polarization is either largely confined to the dielectric coating or extends substantially into the surrounding medium. The dependence on mode order arises because higher-order modes possess more localized field distributions and therefore probe a smaller fraction of the external dielectric environment.

Nonconcentricity accelerates this interpolation, so that the environment-dominated limit is reached already as the interfaces approach the intersection condition $R_0=R_1-\Delta x_1$.

To analyze the optical response, we evaluate the absorption efficiency using \cref{app-eq:Q_abs_noncon_via_B_e_n} together with the corresponding scattered-field coefficients. The results are presented as a function of the actual radius ratio for fixed $\Delta x_1=0.25R_0$ (see \cref{fig:core_shell_disp_rel_Q_abs}~B) and as a function of displacement for fixed $R_1=2R_0$ (see \cref{fig:core_shell_disp_rel_Q_abs}~C).

For small radius ratios, the structure approaches a cylinder with an infinitely thick coating. In this limit, nonconcentricity has only a minor influence on the optical response, and the dipolar resonance dominates the spectrum. As the radius ratio increases, additional multipolar resonances become optically accessible, with each resonance approaching the environment-dominated limit $\omega=\omega_{\rm p}/\sqrt{1+\epsilon_2}$. Beyond the spectral shifts reported in Ref.~\cite{Zhang13}, we find that the direction of the resonance shift depends on the relative dielectric properties of the shell and the surrounding medium.

The scattered potentials shown in the insets of \cref{fig:core_shell_disp_rel_Q_abs}~B display the expected multipolar symmetry of the induced surface charges. At the same time, the corresponding field distributions become increasingly concentrated in and around the thinnest part of the shell. A similar trend is observed when the displacement is increased while keeping $R_1=2R_0$ fixed (\cref{fig:core_shell_disp_rel_Q_abs}~C). In particular, the dipolar near field bends toward the thinner section of the coating eventually only residing in the environment while the field gradually weakens inside the core, signaling a transition toward polarization dominated by the surrounding environment.

In the intersection limit, the scattered-field amplitudes vanish [see \cref{app-eq:B_e_n_limit_a_to_zero}], indicating that localized surface plasmons can no longer be efficiently excited. As discussed in Refs.~\cite{Aubry_2010_PRB,Aubry_2010_NanoLett}, the resulting crescent geometry instead supports propagating surface plasmon modes, which require a different coordinate transformation for their description.

We next consider the dielectric-core--metal-shell configuration ($\epsilon_0=\epsilon_{\mathrm{d}}$ and $\epsilon_1=\epsilon(\omega)$). The corresponding dispersion relation becomes
\begin{align}
	\epsilon^2(\omega)
	+ \epsilon(\omega) (\epsilon_0 + \epsilon_2) \eta_n^{(0,1)}
	+ \epsilon_0\epsilon_2
	&  =
	0,
\end{align}
with
\begin{align}
	\eta_n^{(b,c)}
	=
	\frac{1 + (\tilde{R}_b/\tilde{R}_c)^{2n}}
	{1 - (\tilde{R}_b/\tilde{R}_c)^{2n}} \ge 1  .
\end{align}

The quadratic dependence on the metallic dielectric function reflects the existence of two coupled plasmon branches for each mode order, corresponding to bonding and antibonding hybridized modes\cite{Nordlander2004,Prodan_2003_hybrid_first,Prodan_2004}. Within the Drude model, these resonances are given by
\begin{align}
	\omega^\pm_n &
	= \frac{\omega_{\rm p}}
	{\sqrt{
			1 + \epsilon_0 \mathcal{S}^\pm_n
			(\tilde{R}_0/\tilde{R}_1;
			\epsilon_0/\epsilon_2)
		}
	},
	\label{eq:noncon_diel_core_met_shell_w_LSP}
\end{align}
with
\begin{align}
	\mathcal{S}^{\pm}_n(\tilde{R}_0/\tilde{R}_1;\epsilon_0 /\epsilon_2)
	= & \frac{1}{2}
	\left(
	\frac{\epsilon_2}{\epsilon_0}
	+1
	\right)
	\eta_n^{(0,1)} \nonumber \\
	& \pm  \sqrt{\frac{1}{4}
		\left(
		\frac{\epsilon_2}
		{\epsilon_0}
		+1
		\right)^2 \left(\eta_n^{(0,1)}\right)^2 - \frac{\epsilon_2}
		{\epsilon_0}
	}
	\label{eq:diel_core_met_shell_H_func_disp_rel}
\end{align}

In the intersection limit, the asymptotic behavior of $\mathcal{S}_n^\pm$ yields $\omega_n^+\rightarrow0$ and $\omega_n^-\rightarrow\omega_{\rm p}$, given
\begin{align}
	\mathcal{S}^{\pm}_n
	(\tilde{R}_0/\tilde{R}_1; 
	\epsilon_0 /\epsilon_2)
	\sim \quad~\quad~\quad~\quad~\quad~\quad~\quad~\quad~  \nonumber \\  
	\left\{
	\begin{matrix}
		''+'': &  \eta_n^{(0,1)} \left\{ 1~ + \phantom{4}\dfrac{\epsilon_2}{\epsilon_0} ~~~~~~~~~~+ \mathcal{O}\left[\left(\eta_n^{(0,1)}\right)^{-2}\right]\right\}\\
		''-'': & \eta_n^{(0,1)} \left\{ 4 \dfrac{\epsilon_0\epsilon_2}{\epsilon_0 + \epsilon_2} \left(\eta_n^{(0,1)}\right)^{-2} + \mathcal{O}\left[\left(\eta_n^{(0,1)}\right)^{-4}\right]\right\}
	\end{matrix}
	\right.
\end{align}
This limiting behavior is fully consistent with the transformation-optics picture, in which the transformed geometry corresponds to a metallic slab embedded between two dielectric half-spaces. As the slab thickness vanishes, the two hybrid modes approach precisely these limiting frequencies\cite{Pendry2012_TO_review}. As apparent in \cref{fig:core_shell_disp_rel_Q_abs}~D, increasing nonconcentricity accelerates the convergence toward these asymptotic limits.

The absorption spectra shown in \cref{fig:core_shell_disp_rel_Q_abs}~E,F follow the same asymptotic trends. Again, the potentials reproduce the expected polar patterns, now present on two conductor-dielectric interfaces. Further, for intermediate radius ratios, the bonding modes exhibit field distributions similar to those of the metal-core--dielectric-shell configuration, although their limiting frequencies differ. In contrast, the antibonding modes are predominantly localized within the metallic shell and therefore experience stronger dissipative losses. As nonconcentricity increases, the field maxima of both branches progressively shift toward the thinnest region of the shell.

Overall, the dipolar resonances give rise to the strongest absorption. Bonding modes generally exhibit weaker absorption than antibonding modes because the latter concentrate the electric field more strongly inside the metallic shell, thereby driving dissipative currents more efficiently.

\begin{figure*}
	\centering
	\includegraphics[scale=.8]{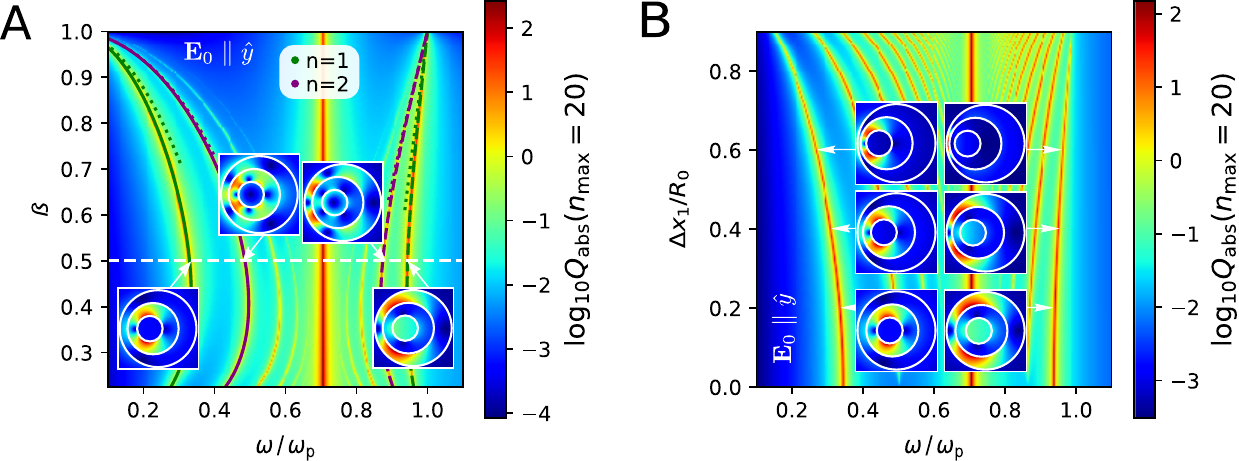}
	\caption{
		\label{fig:abs_eff_tube_wire}
		Absorption efficiency of a tube-wire cavity with core radius $R_0=5\,\mathrm{nm}$, outer radius $R_2=15\,\mathrm{nm}$, shell unit thickness $\Delta R=10\,\mathrm{nm}$, and dielectric environment $\epsilon_{\mathrm{d}}=1$ as a function of frequency.
		Panel A: Influence of the relative shell thickness parameter $\beta$ for a fixed displacement $\Delta x_1=0.25R_0$. The modified dipolar tube-wire cavity resonances are shown for dipolar (green) and quadrupolar (purple) mode orders, with bonding and antibonding branches represented by solid and dashed lines, respectively. The corresponding resonances of the isolated tube structures are indicated by dotted lines in the respective colors. Insets show the electric-field modulus for an external polarization $\mathbf{E_0}\parallel\hat{y}$ at $\beta=0.5$.
		Panel B: Influence of shell displacement $\Delta x_1$ for a fixed relative shell thickness $\beta=0.5$. Insets display the electric-field modulus at the dipolar bonding and antibonding resonances for $\Delta x_1\in\{0.2,0.4,0.6\}\,\mathrm{nm}$ and an external polarization $\mathbf{E_0}\parallel\hat{y}$. 
	}
\end{figure*}

\section{\label{sec:multi_ring_systems} Treatment of core-multi-shell nanowires}

As an example of a multi-shell nanowire, we consider a cross section consisting of a circular core with radius $R_0$ and dielectric response $\epsilon_0$, surrounded by $N$ shell units. Each unit comprises two shells with radial thicknesses $\beta\Delta R$ and $(1-\beta)\Delta R$ [$0\le\beta\le1$], as viewed from a concentric perspective, and dielectric responses $\epsilon_I$ and $\epsilon_{II}$, respectively. The overall size of the structure is determined by the outermost radius $R_{2N}$, such that the radial thickness of each shell unit is fixed by
$\Delta R=\frac{R_{2N}-R_0}{N}$.

The general quasistatic solution for the electrostatic potential, obtained using a transfer-matrix approach, is presented in \cref{app-sec:potential_multi_ring}. In the following, we focus on the representative case in which the core and all even-numbered shells are metallic ($\epsilon_0=\epsilon_{I}=\epsilon(\omega)$), with the dielectric function described by the Drude model using parameters for Ag from Ref.~\cite{PhysRevB.6.4370}. The odd-numbered shells are dielectric with a constant permittivity $\epsilon\subdiel$, chosen to be equal to that of the surrounding medium, i.e., $\epsilon_{II}=\epsilon_{2N+1}=\epsilon\subdiel$. A schematic of the corresponding concentric geometry is shown in \cref{fig:Mie_sketch_con_bulls_eye}. To allow for nonconcentric structures, the outer boundary of the $\alpha$th layer is defined by a circle with center coordinate $x_\alpha$, displaced relative to the inner boundary according to $x_{\alpha-1}=x_\alpha-\Delta x_\alpha$.

\subsection{\label{subsec:tube_wire_cavity}Tube-wire cavity ($N=1$)}

To gain insight into the role of the relative shell thickness within a shell unit, we first consider the simplest example consisting of a single shell unit. This geometry, known as the tube--wire cavity\cite{YuLuo_2012}, has previously been investigated in Refs.~\cite{Luo2013,YuLuo_2012}, where the corresponding Laplace equation was solved analytically.

Since the overall shell-unit thickness $\Delta R$ is fixed, the parameter $\beta$ provides a convenient means of tuning the geometry. Varying $\beta$ changes the position of the inner shell boundary relative to the metallic core and therefore controls the coupling between the core and the metallic shell. At the same time, it modifies the shell thickness and thus the hybridization between the inner and outer shell interfaces. The maximum attainable core--shell separation is determined by the outer radius of the structure.

For $N=1$ with $\epsilon_{\rm I}=\epsilon_3:=\epsilon\subdiel$ and $\epsilon_0=\epsilon_{\rm II}:=\epsilon(\omega)$, the scattered-field coefficient obtained from \cref{app-eq:multi_ring_B_e_n} reads
\begin{align}
	B^{(3)}_n 
	&= 
	-\tilde{E}_0 \frac{G_{n,21}}{G_{n,11}}
	\label{eq:sct_field_coeff_tube_wire}
\end{align}
with $\tilde{E}_0$ given in \cref{app-eq:E_0_exp_coeffs} and 
\begin{align}
	\frac{G_{n,21}}{G_{n,11}} 
	&= 
	-\frac{c_-(\omega)}{c_+(\omega)} 
	\frac{   \Big[ [c_+(\omega)]^2 
		\mu^{(1,2)}_{-,n}
		+ g^{(1,2)}_{-,n}(\omega) (\tilde{R}_0^{2n}/\tilde{R}_1^{2n})  
		\Big] \tilde{R}_{2}^{2n}
	}
	{
		g^{(1,2)}_{+,n}(\omega) 
		- [c_-(\omega)]^2   (\tilde{R}_1^{-2n}-\tilde{R}_2^{-2n}) \tilde{R}_0^{2n}
	}
	\label{eq:G_mat_coeff_frac_tube_wire}
\end{align}
where
\begin{align}
	c_\pm(\omega) 
	& = 
	\epsilon(\omega) \pm \epsilon_{\rm d} \\
	g^{(b,c)}_{\pm,n}(\omega) 
	& = 
	\pm\left[ \epsilon^2(\omega) + \epsilon^2_{\rm d} \right] 
	\mu^{(b,c)}_{-,n} + 2\epsilon_{\rm d} \epsilon(\omega) 
	\mu^{(b,c)}_{+,n},
\end{align}
with
\begin{align}
	\mu^{(b,c)}_{\pm,n} = 1 \pm \frac{\tilde{R}^{2n}_b}{\tilde{R}^{2n}_{c}}.
\end{align}

The resonance frequencies follow from the zeros of the denominator in \cref{eq:G_mat_coeff_frac_tube_wire}. The denominator factorizes into two contributions. The first corresponds to the isolated metallic cylinder, $c_+(\omega)=0$, whose resonance also appears in Refs.~\cite{Luo2013,YuLuo_2012}. The second yields the modified shell resonances. These reduce to the resonances of an isolated metallic shell, $g^{(1,2)}_{+,n}(\omega)=0$, in the absence of core--shell coupling, while the remaining term accounts for the interaction between the metallic core and shell. As expected, this coupling depends on all three radii.

For fixed $\Delta x_1$, the parameter $\beta$ possesses a lower bound determined by the intersection of the inner two interfaces, $R_1(\beta)\rightarrow R_0+\Delta x_1$, corresponding to a two-crescent-shell geometry. In this limit, the focal parameter vanishes and the circle centers coincide with their radii according to \cref{app-eq:focal_param_via_one_circle,app-eq:x_c_alpha_via_radii_and_cent_displ}. Consequently, the scattered-field coefficient reduces to the isolated-cylinder pole, while the absorption efficiency obtained from \cref{app-eq:Q_abs_noncon_via_B_e_n} vanishes. In contrast, no analogous upper bound exists for $\beta<1$. As $\beta\rightarrow1$, the metal shell becomes concentric [$\Delta x_2\rightarrow0$], and the optical response continuously approaches that of the corresponding concentric tube.

The modified shell resonances split into low-frequency (bonding) and high-frequency (antibonding) branches, analogous to the plasmon hybridization of an isolated metallic shell with dielectric core. Their resonance frequencies are determined by
\begin{align}
	\mathcal{F}^\pm_n(\tilde{R}_0,&\tilde{R}_1,\tilde{R}_2)
	= \frac{1}{\mu^{(0,1)}_{-,n}} 
	\left( 
	\eta_n^{(1,2)} 
	+ \frac{\tilde{R}^{2n}_0}{\tilde{R}^{2n}_1}
	\right)                    \nonumber \\ 
	\pm & \sqrt{
		\left[ \frac{1}{\mu^{(0,1)}_{-,n}} 
		\left( 
		\eta_n^{(1,2)} 
		+ \frac{\tilde{R}^{2n}_0}{\tilde{R}^{2n}_1}
		\right)
		\right]^2 - 1	
	}
	\label{eq:tube_wire_F_func_disp_rel}
\end{align}
Both branches recover the isolated-shell limits as $\beta\rightarrow1$. In the opposite limit, $\beta=\beta_{\rm min}$, the bonding modes soften towards zero frequency, whereas the antibonding modes approach the bulk plasma frequency.

These limiting behaviors are reflected in the absorption spectra shown in \cref{fig:abs_eff_tube_wire}. \Cref{fig:abs_eff_tube_wire}~A displays the absorption efficiency for fixed $R_0=5$ nm, $R_2=15$ nm, $\Delta R=10$ nm, $\Delta x_1=1.25$ nm ($0.25R_0$), and $\epsilon\subdiel=1$, while varying $\beta$. The cylindrical resonance is present throughout the entire parameter range, although its strength decreases rapidly as the intersection limit is approached. The bonding and antibonding branches evolve smoothly between their limiting frequencies, with the spectral separation becoming smallest at intermediate values of $\beta$. For sufficiently large $\beta$, the shell resonances closely resemble those of an isolated metallic tube, particularly for higher-order modes. In contrast, as the intersection limit is approached, the enhanced sensitivity to nonconcentricity gives rise to an increasing number of optically active resonances.

The field distributions shown in the insets further illustrate the underlying plasmon hybridization. The antibonding modes are predominantly localized within the metallic shell, explaining their close correspondence with the isolated-shell resonances. In contrast, the bonding modes concentrate the field within the dielectric gap between core and shell, making them substantially more sensitive to core--shell coupling. Moreover, higher-order modes exhibit stronger confinement to the metallic shell and therefore more closely reproduce the isolated-shell resonance frequencies.

\Cref{fig:abs_eff_tube_wire}~B shows the effect of increasing the nonconcentricity while keeping $\beta=0.5$ fixed. Increasing the displacement broadens the spectral range and activates higher-order resonances, while each hybridized branch approaches its respective crescent-limit frequency. In the concentric limit, only the dipolar cylinder resonance and the dipolar modified-shell resonances remain optically active. As anticipated from \cref{eq:G_mat_coeff_frac_tube_wire}, the cylindrical resonance frequency itself remains unaffected by nonconcentricity.

The corresponding near-field distributions reveal that increasing nonconcentricity progressively concentrates the electromagnetic field within the thinnest section of the respective shell. For the bonding modes, this localization is accompanied by an increasing field spill-out into the surrounding dielectric.

Although the cylindrical resonance remains spectrally unchanged, its scattering amplitude acquires a geometry-dependent prefactor. Setting $c_+(\omega)=0$ in the second factor of \cref{eq:G_mat_coeff_frac_tube_wire} and substituting the result into \cref{eq:sct_field_coeff_tube_wire} yields
\begin{align}
	B^{(3)}_n & \to \tilde{E}_0 \frac{c_-(\omega)}{c_+(\omega)} \tilde{R}^{2n}_0~\times \\
	& \quad ~~~  \times
	\left[
	\frac{(\tilde{R}^{2n}_2 / \tilde{R}^{2n}_1)}
	{(\tilde{R}_1^{2n} / \tilde{R}^{2n}_2) 
		+ (\tilde{R}_0^{2n} / \tilde{R}_1^{2n} - \tilde{R}_0^{2n} / \tilde{R}^{2n}_2) } \right]
	\nonumber 
\end{align}
where the term in brackets quantifies the deviation from the scattering coefficient of an isolated metallic cylinder in the concentric limit.

\begin{figure*}[t]
	\centering
	\includegraphics[scale=.5]{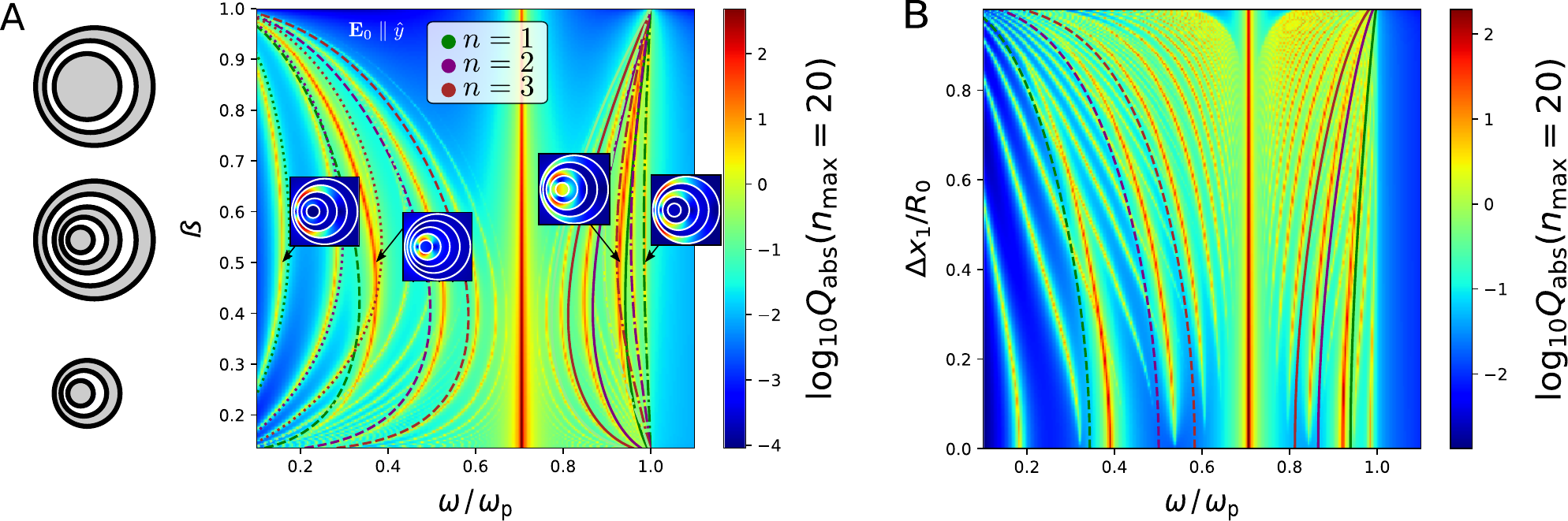}
	\caption{
		\label{fig:ncon_bulls_eye_Q_abs_increase_peaks}
		Absorption efficiency of a bipolar bull's eye wire consisting of two shell units, with core radius $R_0=5\,\mathrm{nm}$, outer radius $R_{2N}=25\,\mathrm{nm}$, and dielectric environment $\epsilon_{\mathrm{d}}=1$. The corresponding geometry is shown in the central sketch of the left subpanel of panel (A).
		Panel A: Dependence on the relative shell thickness parameter $\beta$ for a fixed displacement $\Delta x_1=1.25\,\mathrm{nm}$ ($0.25R_0$). The resonance frequencies of bonding (dashed lines) and antibonding (solid lines) hybridized plasmons for different mode orders are shown for the tube-wire cavity obtained by removing the outer shell unit (lower sketch in the left subpanel). For comparison, the bonding (dotted lines) and antibonding (dash-dotted lines) resonance frequencies of the complementary tube-wire cavity obtained by replacing the innermost dielectric gap with metal are also shown (upper sketch in the left subpanel). Insets display the modulus of the electric field at resonance for $\beta=0.5$ and external polarization $\mathbf{E_0}\parallel\hat{y}$.
		Panel B: Dependence on the displacement $\Delta x_1$ for a fixed relative shell thickness parameter $\beta=0.5$. The resonance frequencies of bonding (dashed lines) and antibonding (solid lines) hybridized plasmons for different mode orders are shown for the tube-wire cavity obtained by removing the outer shell unit.
        }	
\end{figure*}

\subsection{\label{subsec:multi_ring_mu_greater_one}Bull's eye nanowires ($N>1$)}

From the analysis of core--single-shell and tube-wire cavity structures, we identified two geometrical parameters that control the number and distribution of resonances in the absorption spectrum: the degree of nonconcentricity, quantified by the displacement $\Delta x_1$, and the relative shell thickness parameter $\beta$, which determines the spectral separation and resolution of the resulting modes. When extending the system to more complex core--multi-shell structures, the number of shell units introduces a third geometrical degree of freedom. For nonconcentric structures whose interfaces follow bipolar coordinate lines, we refer to the resulting geometry as a bipolar bull's eye wire.

As a first step, we consider a bipolar bull's eye wire consisting of two shell units. In \cref{fig:ncon_bulls_eye_Q_abs_increase_peaks}~A, we show the absorption efficiency as a function of $\beta$ for $R_0=5\,\mathrm{nm}$, $R_{2N}=25\,\mathrm{nm}$, $\Delta x_1=1.25\,\mathrm{nm}=0.25R_0$, and $\epsilon_{\mathrm{d}}=1$ [central geometry sketch in the left subpanel]. The spectrum exhibits a set of dominant resonances located around the central cylindrical surface plasmon frequency $\omega=\omega_{\rm p}/\sqrt{2}$. These resonances are approximately reproduced by the bonding and antibonding modes of a tube-wire cavity obtained by removing the outer shell unit [lower sketch in the left subpanel]. In addition, weaker resonances extend the spectrum towards lower and higher frequencies. These modes exhibit crossings with the dominant resonance group and are approximately described by the tube-wire cavity obtained by replacing the innermost dielectric gap by metal [upper sketch in the left subpanel].

The correspondence with the tube-wire cavity modes indicates that the plasmonic response of the bull's eye wire can be interpreted within a hybridization picture, where only a subset of neighboring conductor--dielectric interfaces contributes significantly to individual resonances. This interpretation is supported by the electric-field distributions shown in the insets. Furthermore, the approximate tube-wire description remains valid in the limiting cases $\beta\rightarrow0$ and $\beta\rightarrow1$. The broadening of the cylindrical resonance for $\beta\rightarrow0$ can be attributed to the increasing metallic fraction of the structure.

A similar behavior is observed when varying the nonconcentric displacement. In \cref{fig:ncon_bulls_eye_Q_abs_increase_peaks}~B, we fix $\beta=0.5$ and increase $\Delta x_1$. Again, the dominant resonances are approximately described by the corresponding tube-wire cavity modes, while the number of accessible resonances increases with increasing nonconcentricity. In the concentric limit, only the dipolar resonances remain optically active. Approaching the intersection limit (crescent-onion geometry), the characteristic resonance frequencies of the individual interfaces are recovered, while the spectrum ultimately reduces towards the cylindrical surface plasmon response.

\begin{figure}[h]
	\centering
	\includegraphics[scale=.75]{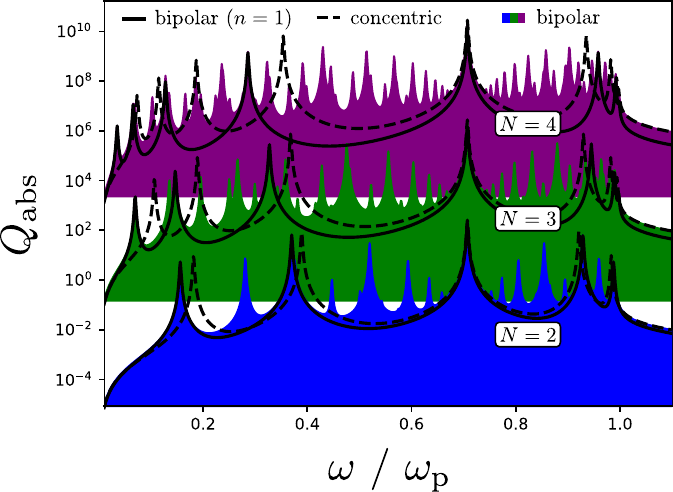}
	\caption{
		\label{fig:bulls_eye_different_unit_cell_amounts}
		Absorption efficiency of bull's eye wires with increasing number of shell units for $R_0=5\,\mathrm{nm}$, $R_{2N}=25\,\mathrm{nm}$, $\beta=0.5$, $\Delta x_1=1.25\,\mathrm{nm}$, and dielectric environment $\epsilon_{\mathrm{d}}=1$ along with a
		comparison of concentric (dashed lines) and bipolar nonconcentric bull's eye wires (shaded regions) for $N\in\{2,3,4\}$ on a logarithmic absorption-efficiency scale. The dipolar contribution of the nonconcentric structures is additionally shown as solid lines. For clarity, the spectra for increasing numbers of shell units are vertically offset by factors of $10^4$. 
}     
\end{figure}

We next investigate the influence of the number of shell unit. As shown in \cref{fig:bulls_eye_different_unit_cell_amounts}, increasing the number of shell units introduces additional resonances even in the concentric limit. Within our analytical description, these modes are predominantly dipolar and originate from the increasing number of metal--dielectric interfaces. Due to the finite spatial extent of the near fields, coupling is strongest between neighboring interfaces. In the bipolar nonconcentric case, optical excitation of higher-order modes introduces a substantially larger number of resonances for increasing shell-unit count. These additional resonances are primarily located between those of the corresponding concentric structures and therefore increase the spectral density more strongly than the overall spectral width.

Relative to the concentric bull's eye wire, the dipolar resonances below (above) $\omega=\omega_{\rm p}/\sqrt{2}$ exhibit redshifts (blueshifts) due to nonconcentricity. Furthermore, their reduced amplitudes indicate weaker coupling to the incident field. Notably, the resonance associated with the cylindrical surface plasmon frequency remains present even when increasing the number of shell units.

\begin{figure} 	
 	\centering
 	\includegraphics[scale=.58]{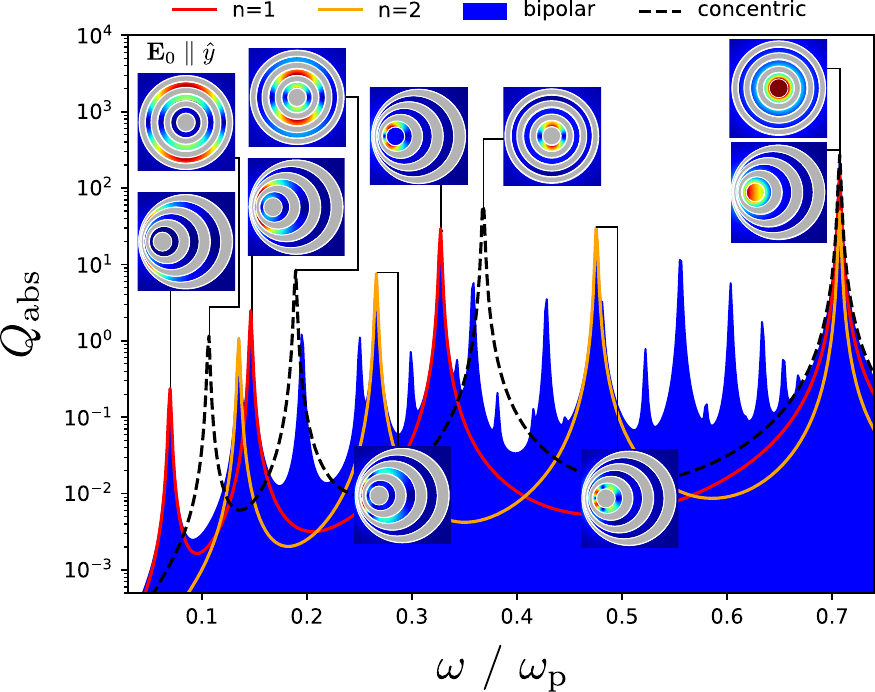}
 	\caption{
 		\label{fig:ncon_bulls_eye_N_3_with_insets}
 		Absorption efficiency in the low-frequency spectral range of a bipolar bull's eye wire with core radius $R_{0}=5\,\mathrm{nm}$, outer radius $R_{2N}=25\,\mathrm{nm}$, relative shell thickness parameter $\beta=0.5$, displacement $\Delta x_1=1.25\,\mathrm{nm}$, and dielectric environment $\epsilon_{\mathrm{d}}=1$. The total spectrum obtained by including multipole orders up to $n_{\mathrm{max}}=10$ is shown as a blue shaded region.  Dipolar and quadrupolar contribution are shown as colored lines and compared with the spectrum of the corresponding concentric bull's eye wire (black dashed line). Insets display the modulus of the scattered electric field at selected resonances for the respective geometries given $\mathbf{E}_0\parallel \hat{y}$. The grey coloring of metallic regions is included as a visual guide
 		apart from the core wire in case of the cylindrical resonance.
 		 } 
\end{figure}

To identify the physical origin of the additional resonances, we analyze the contribution of individual mode orders for a three-shell-unit bipolar bull's eye wire in \cref{fig:ncon_bulls_eye_N_3_with_insets}. We focus on the lower-frequency spectral region, where the corresponding modes exhibit field maxima inside dielectric gaps and are therefore particularly relevant for near-field-driven processes such as photopolymerization. Within the present approximation, no hybridization between different angular mode orders is observed. In contrast to systems such as closely spaced spherical dimers, where Coulomb coupling between different single-particle multipole orders leads to mixed hybridized modes\cite{Nordlander2004}, each resonance here can be assigned predominantly to an individual mode order.

An inspection of the electric-field distributions for both concentric and nonconcentric bull's eye wires reveals that the plasmonic hot spots are localized within individual dielectric gaps. The specific gap hosting the maximum field depends on the resonance frequency, while nonconcentricity shifts the hot spots towards the thinner shell regions. For a given geometry, decreasing the resonance frequency corresponds to a systematic movement of the field localization towards the outer dielectric shells. This behavior is also observed within individual higher-order modal contributions.

For example, quadrupolar resonances appear at higher frequencies than their dipolar counterparts while exhibiting analogous localization trends. A quadrupolar mode localized in the innermost dielectric shell occurs at $\omega\approx0.476\,\omega_{\rm p}$, compared with the corresponding dipolar resonance at $\omega\approx0.328\,\omega_{\rm p}$. Similarly, localization within the second dielectric shell is obtained for a quadrupolar mode at $\omega\approx0.265\,\omega_{\rm p}$, compared with the dipolar resonance at $\omega\approx0.146\,\omega_{\rm p}$. The two-lobed structure of the quadrupolar field distribution becomes particularly pronounced within the thin shell regions.

The observed field localization further supports the interpretation that individual bull's eye resonances can be associated with hybridization between neighboring interfaces surrounding a given dielectric gap. For instance, the highest-frequency resonance of a given mode order below $\omega_{\rm p}/\sqrt{2}$, corresponding to localization within the innermost dielectric gap, is approximately reproduced both in resonance frequency and field distribution by the corresponding tube-wire cavity approximation. Thus, the bull's eye wire may be interpreted as a collection of coupled tube-wire cavities. Since the considered tube-wire resonances are predominantly bonding modes, particularly for large $\beta$, the low-frequency resonances of the bull's eye structures can likewise be approximately assigned a bonding character.

 \subsection{\label{subsec:multi_ring_compare_doppler} Comparison of bipolar and doppler bull's eye nanowires}
 
 Having established the geometrical parameters governing the spectral response of concentric and bipolar bull's eye wires, we now compare the latter with a geometry that mimics the horizontal cross section of a Doppler grating~\cite{See2017}. The Doppler bull's eye wire is constructed by retaining the same radial discretization and setting $\beta=0.5$. In contrast to the bipolar geometry, where the shell interfaces coincide with bipolar coordinate lines, the Doppler geometry is defined by a uniform displacement of all interfaces, $\Delta x^{\mathrm{dop}}_\alpha=\mathrm{const.}$ ($\alpha\in\{1,\dots,2N\}$). Unlike Ref.~\cite{See2017}, which considered grating disks, we focus here on infinitely extended wires. The cross-sectional geometry is defined by $R_0=5\,\mathrm{nm}$, $R_{2N}=20\,\mathrm{nm}$, and $\Delta R=5\,\mathrm{nm}$. Interpreting nonconcentricity as a fabrication-induced deviation, we consider the relatively small displacement $\Delta x_1=0.3\,\mathrm{nm}$. The spectra of the concentric, bipolar, and Doppler bull's eye wires are obtained directly using the DGTD method (see \cref{app-sec:num_method}).
 
\begin{figure*}[t]
 	\centering
 	\includegraphics[scale=.75]{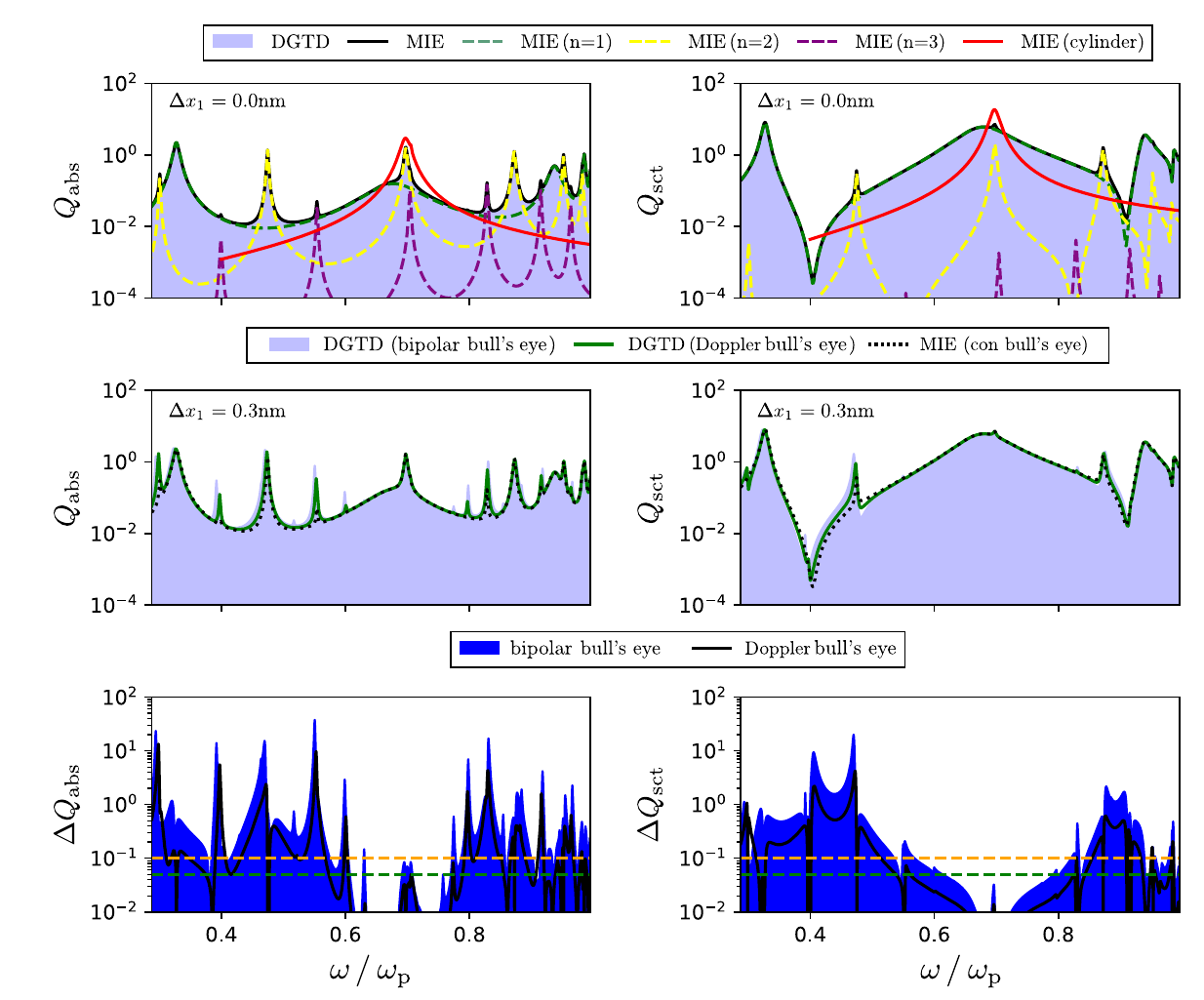}
 	 	\caption{
 		\label{fig:bulls_eye_vs_doppler_spectra}
 		Absorption (left column) and scattering (right column) efficiencies of concentric and nonconcentric bull's eye wires consisting of three shell units. The structures have core radius $R_0=5\,\mathrm{nm}$, outer radius $R_{2N}=20\,\mathrm{nm}$, and relative shell thickness parameter $\beta=0.5$.
 		Top row: Concentric bull's eye wire. The results obtained from cylindrical Mie theory are compared with discontinuous Galerkin time-domain (DGTD) simulations. The individual multipolar contributions from dipolar to hexapolar order extracted from the Mie solution are shown according to the legend.
 		Middle row: Comparison of the absorption and scattering efficiencies of the concentric bull's eye wire with those of bipolar and Doppler bull's eye wires for a fixed displacement $\Delta x_1=0.3\,\mathrm{nm}$.
 		Bottom row: Relative frequency-dependent difference between the simulated spectra of the bipolar and Doppler bull's eye wires and the Mie-theory spectrum of the concentric bull's eye wire. The orange and green dashed lines indicate relative deviations of $10\%$ and $5\%$, respectively. 
 		}	
 \end{figure*}

 Considering the upper row of \cref{fig:bulls_eye_vs_doppler_spectra}, we first verify that the absorption and scattering efficiencies of the concentric bull's eye wire obtained from DGTD simulations are accurately reproduced by the cylindrical Mie-theory calculation described in \cref{sec:Mie_theory}. The analytical solution further enables decomposition into individual multipole orders, facilitating mode assignment. The spectrum is found to be dominated by dipolar through hexapolar contributions, with decreasing weight for increasing multipole order. For comparison, the response of an isolated cylindrical wire with radius $R_{0}=5\,\mathrm{nm}$ is also shown. In the absorption spectrum, the dipolar and quadrupolar resonances of the bull's eye wire around $\omega_{\rm p} / \sqrt{2}$ are redshifted towards the corresponding cylindrical-wire resonances. At the same time, the dipolar bull's eye resonance exhibits a substantially larger spectral broadening, resulting in a modified relative contribution of dipolar and quadrupolar modes compared with the isolated cylinder. A similar broadening is observed in the scattering spectrum, where the quadrupolar contribution of the isolated cylinder is, however, negligible.
 
 The second and third rows of \cref{fig:bulls_eye_vs_doppler_spectra} compare the absorption and scattering efficiencies of the bipolar and Doppler bull's eye wires with those of the concentric reference structure. The smallest deviations occur near the resonances of the isolated cylindrical wire. In this spectral region, the relative deviations of the absorption (scattering) efficiency remain below $10\%$ ($5\%$) for the bipolar bull's eye wire and below $5\%$ ($5\%$) for the Doppler geometry. Larger deviations emerge towards the spectral edges, where the nonconcentric structures exhibit resonance shifts and amplitude variations. Interestingly, the resonance shifts are more pronounced for the nonconcentric bull's eye geometries than for the concentric structure.

\section{\label{sec:conclusion_and_outlook}Conclusions and Outlook}
 
In this work, we have investigated the influence of shell nonconcentricity on the quasistatic optical response of core--single-shell and bull's eye core--multi-shell nanowires. Our analysis was based on the absorption efficiency and the spatial field distributions of selected localized surface plasmon (LSP) modes.

The core--single-shell nanowires provide the simplest geometry for elucidating the influence of nonconcentricity. In particular, increasing shell nonconcentricness leads to a finite coupling of the incident field to higher-order LSP modes that remain dark in the concentric limit. Building upon these findings, we considered bull's eye nanowires consisting of a metallic core surrounded by a finite number of shell units, each comprising a dielectric shell followed by a metallic shell. To disentangle the influence of the individual geometrical parameters, we first investigated the tube-wire cavity, corresponding to a single shell unit. The analytically derived dispersion relations reveal the distinct roles of shell-interface nonconcentricity and the relative dielectric-to-metal thickness within each unit. Increasing the number of units introduces an additional geometrical degree of freedom, giving rise to increasingly complex hybridized plasmon spectra.

Even in the concentric limit, increasing the number of metal--dielectric interfaces broadens the plasmonic spectrum through the emergence of additional hybridized modes. Shell nonconcentricity further increases the spectral density by enabling optical excitation of higher-order resonances. Simultaneously, the corresponding near fields become progressively concentrated around the thinnest shell regions. While the resulting broad spectral response is attractive for light-harvesting applications, the resonance amplitudes eventually decrease as the geometry approaches the crescent limit. In this regime, propagating surface plasmon modes have been studied employing a different conformal transformation~\cite{Aubry_2010_PRB,Aubry_2010_NanoLett}.

Perhaps the most intriguing result is the existence of spectrally well-separated LSP modes whose near fields are selectively localized within individual dielectric gaps of the bull's eye structure. Such spatial selectivity may be exploited for localized photopolymerization, where the polymer growth follows the plasmonic hot spots, or for the selective excitation of optically active constituents embedded within chosen dielectric layers. In particular, different dielectric gaps could be functionalized with distinct dye species whose absorption spectra are matched to individual LSP resonances, thereby enabling spectrally addressable plasmon--emitter coupling as an extension to, e.g Refs.~\cite{Stete2017,SteteHeurQuantumModell}.

For sufficiently strong, yet not extreme, shell nonconcentricity, the highly localized near fields may further provide a promising platform for coupling to few or even single emitters~\cite{Kewes2018}. At the same time, the pronounced angular anisotropy of the plasmonic near fields should be taken into account when designing photopolymerization processes. A particularly attractive direction would therefore be the realization of molecularly imprinted polymer shells~\cite{Khitous2023PlasmonInducedPO,Khitous2023} supported by concentric or nonconcentric bull's eye nanowires. Depending on the application, monochromatic illumination could selectively activate individual dielectric layers, whereas polychromatic excitation may simultaneously address multiple layers.

Several extensions of the present work remain to be explored. From a theoretical perspective, the influence of nonlocal response and the finite height of realistic nanowires should be assessed. Furthermore, a direct comparison between bull's eye nanodisks derived from bipolar coordinate lines and Doppler-grating-inspired nanodisks would establish a closer connection to experimentally realized structures. Such a study would clarify whether Doppler-grating geometries can support localized surface plasmon functionalities in addition to the propagating surface plasmon polaritons that have been extensively investigated for micrometer-scale Doppler gratings in applications such as gas sensing, color filtering, and related photonic devices.

\begin{acknowledgements}
The authors gratefully acknowledge financial support by
the German Research Foundation (DFG) within the frame-
work of the Collaborative Research Center 1375 ’Nonlinear
Optics down to Atomic Scales (NOA)’ (Project ID 398816777
- Projects A6 and C1) and valuable discussions with Dr.
Francesco Intravaia on the transfer-matrix method and the
bipolar coordinates.
\end{acknowledgements}

\appendix

\section{\label{app-sec:potential_ring_structures}Potentials of the core-single/multi-shell nanowires}

Within the manuscript, we have treated a multitude of optical properties based on the potentials of selected structures that solve Laplace equation. In the following, we shall derive the potentials.

\subsection{Expansion of Laplace equation}
Based on vertical homogeneity of the scatterer and source,
we restrict to two dimensions, where conformal transformations from cartesian to certain coordinate systems exist in which the
Laplace equation separates\cite{morse1953methods}. Such systems possess identical scaling factors 
$h(s, w)$, such that
\begin{align}
0 = \nabla \cdot \nabla V = \frac{1}{h^2(s, w)} \left[ \partial_{s}^2 + \partial_{w}^2\right] V(s,w) 
\label{eq:expansion_laplace_eq_conforma_trafo}
\end{align} 
The solution is considered to be a piecewise concatenation of the potentials of each individual medium. Further, the latter are labeled by an index $\alpha$, where the core is marked by $\alpha{=}0$ and bounded by the radius $R_0$ centered at $y=0$ and $x=x_0$ with permittivity $\epsilon_0$. Each shell (labeled $\alpha{\ge}1$ with permittivity $\epsilon_\alpha$) is bounded by circles of radii $R_\alpha$ and $R_{\alpha-1}({<}R_\alpha)$ and with center coordinates $x_{\alpha}$ and $x_{\alpha-1}({<}x_{\alpha})$ defining a horizontal displacement of $\Delta x_{\alpha} = x_{\alpha}-x_{\alpha-1}$. Such circles are aligned with appropriately selected coordinate-lines, such that the surface-normal and -tangential unit vectors $\hat{n}$ and $\hat{t}$ coincide with the
unit vectors $\hat{s}$ and $\hat{w}$ of the chosen coordinate system (apart from, possibly, a global sign). In our notation, the coordinate $s$ traverses through the media with interfaces parametrized by given values, while $w\in[0,2\pi)$ allows to traverse the interface. The potential will always be $2\pi$-periodic in $w$ which introduces the azimuthal 
constant $m>0$, which fixes the multipolar order of the corresponding field contribution.
The particular coordinate systems are introduced in the following. 

\begin{figure*}
	\centering
	\includegraphics[scale=.75]{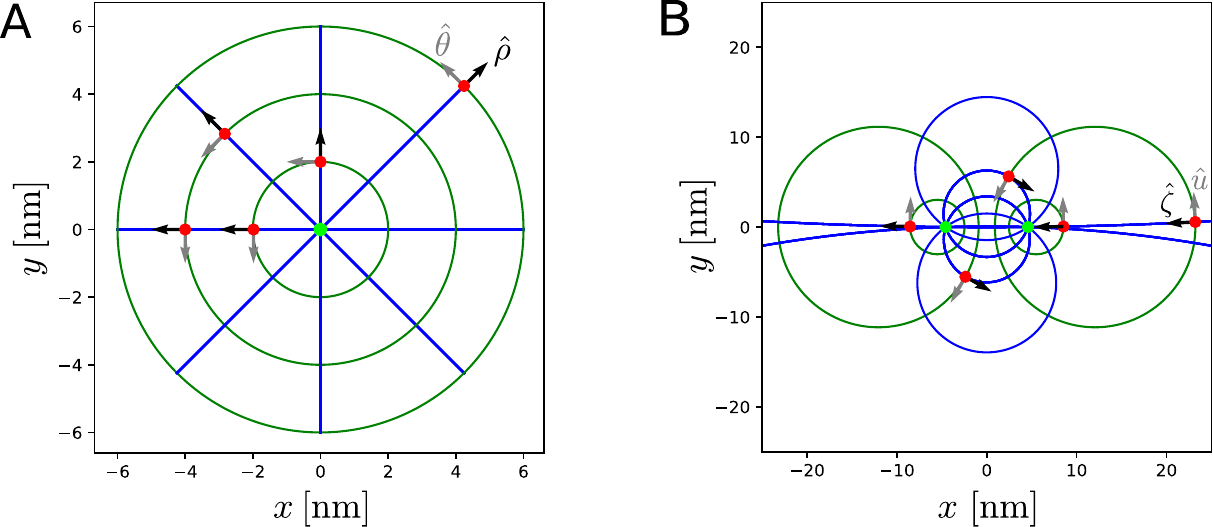}
	\caption{
		\label{fig:conf_coord_sys}
		Coordinate systems in the Cartesian frame: log-polar (Panel A) and bipolar coordinates (Panel B). The coordinate lines corresponding to the $s$ and $w$ coordinates are shown in green and blue, respectively, with the coordinate poles indicated by green dots. The local unit vectors $\hat{s}$ (black arrows) and $\hat{w}$ (grey arrows) vary spatially when expressed in the Cartesian frame but remain orthogonal at every point. In panel B, the branch cut along $x\in(-\infty,-a]\cup[a,\infty)$ and $y=0$ is implicit.
	}
\end{figure*}

\subsection{Log-polar coordinates}
For the wires with concentric shells, we apply the logarithmic polar coordinates, 
where $s$ is given by the logarithm of the radius, i.e., by $\rho = \ln \sqrt{x^2+y^2} \in \mathbb{R}$, while $w$ is embodied by the polar angle $\theta$. The $\rho$-coordinate lines are concentric circles with respect to a
single pole at $x=y=0$ with radii $\tilde{R}(\rho)=\exp(\rho)$ being equal to the metric factors. 
The linearly-polarized electric field transforms according to
\begin{align}
V_0(\rho, \theta) = -E_0 \exp(\rho) \left[ \cos(\tau) \cos (\theta) + \sin(\tau) \sin(\theta) \right]
\label{app-eq:V_0_log_pol}
\end{align} 
where $\gamma=\angle(\vec{E_0}, \hat{x})$ is the angle between the $x$-axis and the polarization of the incident field counted
in counter-clockwise orientation. Only dipolar modes
can be excited on the concentric-shell wires. The unit vectors read
\begin{align}
 \hat{\rho}   &=   \phantom{-}\cos(\theta) \hat{x} 
                 + \sin(\theta) \hat{y} \\
 \hat{\theta} &=            - \sin(\theta) \hat{x} 
                 + \cos(\theta) \hat{y}  
\end{align}
and fullfill the cross product
\begin{align}
 \hat{\rho}   \times \hat{\theta} & = \hat{z}      
\end{align}
with the other products following from the triple product expansion and orthogonality of the unit vectors.
For a depiction of the coordinates lines and unit vectors, see \cref{fig:conf_coord_sys}~A.
\subsection{Bipolar coordinates}
For the wires with nonconcentric shells, we employ the bipolar coordinate system.
For better distinction from the concentric pendants, $s$ is then represented by some $\zeta\in \mathbb{R}$ and $w$ by $u$. 
The transformation to cartesian coordinates reads\footnote{We note, that our convention regarding the ''angular'' variable $u$ differs slightly from the variable $\theta$ used in Ref.~\cite{morse1953methods}. We place a 
	branch cut at $y=0$ and $x\in(-\infty, -a]\cup[a, \infty)$ in line with Ref.~\cite{LuchtBipolar} 
	differing from the choice of $y=0$ and $x\in[-a, a]$ 
	within Ref.~\cite{morse1953methods}. While in both conventions, the angular
	coordinate lines all meet at $y=0$ and $x=\pm a$, $u$ grows counter-clockwise 
	while $\theta$ grows clockwise. This leads to the mappings 
	$\cos(m\theta)= (-1)^m \cos(m u)$ and $\sin(m\theta) = (-1)^{m+1}\sin(m u)$.}
\begin{align}
x(\zeta,u) &= h(\zeta,u) \sinh(\zeta) 
\label{eq:x_of_u_zeta}\\
y(\zeta,u) &= h(\zeta,u) \sin(u) 
\label{eq:y_of_u_zeta}\\
h(\zeta,u) &= \frac{a}{\cosh(\zeta) - \cos(u)}
\label{eq:scale_fac_u_zeta}
\end{align}
given a focal parameter $a$ defining two focus points
at $y=0$ and $x=\pm a$. Close to these poles, $\zeta$ and 
$u$ assume the roles of the log-polar radial and angular coordinates,
respectively.
The different media are delineated by $\zeta$-lines which define
nonconcentric circles of radius $R(\zeta) = a / |\sinh(\zeta)|$ and
center at $(y=0, x= a \coth(\zeta))$ which coalesce at the focal points of the respective
halfspace as $\zeta\to \pm \infty$. 
Already, the focal parameter (and therefore the coordinate system) is fully determined by a single circle, as 
\begin{align}
a^2 = x^2(\zeta) - R^2(\zeta) = \textrm{const.}
\label{app-eq:focal_param_via_one_circle}
\end{align}
Due to the translation invariance of experiments, it is more meaningful to introduce the displacement $\Delta x_\alpha{=} x(\zeta_{\alpha}){-}x(\zeta_{\alpha-1})$ of the circles bounding each shell medium with label $\alpha>1$ rather than their absolute positions. This defines the center of the shell's inner circle as
\begin{align}
x_{\alpha-1} & = \frac{1}{2\Delta x_\alpha} \left[ R^2_{\alpha}- R^2_{\alpha-1} - \Delta x^2_{\alpha}\right] ~~~, ~~~ \alpha\ge 1
\label{app-eq:x_c_alpha_via_radii_and_cent_displ}
\end{align}
where $R_\alpha = R(\zeta_\alpha)$ and $x_\alpha = x(\zeta_\alpha)$.
Note, that by \cref{app-eq:focal_param_via_one_circle} and \cref{app-eq:x_c_alpha_via_radii_and_cent_displ}, the focal parameter is specified by definition of a single shell via its boundaries with radii $(R_{\alpha}, R_{\alpha-1})$ and center displacement $\Delta x_\alpha$. For each new circle which marks the onset of another shell (or the environment) either the displacement or the radius can be chosen freely. So,  when going beyond the core-single-shell wire, we cannot parametrize arbitrary nonconcentric
arrangements with the bipolar coordinates. 
Generally, a particular limit is obtained from \cref{app-eq:x_c_alpha_via_radii_and_cent_displ} following  
\begin{align}
R_{\alpha} = R_{\alpha-1} + \Delta x_{\alpha}
\Leftrightarrow
x_\alpha = R_\alpha  
\label{eq:crescent_limit}
\end{align}
For the core-single/multi-shell wires, this is the crescent-shape limit.

Within the bipolar coordinates, the potential of the uniform field reads
\begin{align}
& V_0(u,\zeta) = -a E_0 \sum\limits_{n=0}^{\infty} \exp(-n|\zeta|) \times  
\label{app-eq:V_0_bipo} \\
&  \quad \times (2- \delta_{n0}) 
       \left[
            \sgn{\zeta} \cos(\tau)\cos(nu) 
          +             \sin(\tau) \sin(nu)
      \right] \nonumber
\end{align}
which can be shown by a Fourier series expansion of \cref{eq:x_of_u_zeta,eq:y_of_u_zeta} 
for $\zeta{\ne}0$. Note, how the transition to a nonconcentric arrangement allows for the excitation of modes beyond the dipolar order. In our studies, we need only restrict to the right half space with 
$\zeta{>}0$, where the signum function equals unity. 

The unit vectors can be expressed in terms of their cartesian counterparts by 
\begin{align}
\hat{\zeta} 
&= \frac{1}{h(\zeta,u)} \left[ \partial_\zeta x(\zeta,u) ~\hat{x} + \partial_\zeta y(\zeta,u) ~\hat{y} \right] \nonumber \\
&=   \frac{1-\cosh(\zeta)\cos(u)}{\cosh(\zeta) - \cos(u)} \hat{x} 
- \frac{\sinh(\zeta)\sin(u)}{\cosh(\zeta) - \cos(u)} ~\hat{y}
\label{eq:def_hat_zeta} \\
\hat{u} 
&= \frac{1}{h(\zeta,u)}  \left[ \partial_u x(\zeta,u) ~\hat{x} + \partial_u y(\zeta,u)~ \hat{y} \right] \nonumber \\
&= -\frac{\sinh(\zeta)\sin(u)}{\cosh(\zeta) - \cos(u)}~ \hat{x}
-\frac{1-\cosh(\zeta)\cos(u)}{\cosh(\zeta) - \cos(u)} ~\hat{y}    
\label{eq:def_hat_u}                 
\end{align}
They obey the cross product
\begin{align}
\hat{u}    \times \hat{\rho} & = \hat{z}    
\label{eq:bipo_hat_u_cross_prod_hat_rho} 
\end{align}
with the other products following from the triple product expansion and orthogonality of the unit vectors.
For a depiction of the coordinates lines and unit vectors, see \cref{fig:conf_coord_sys}~B. We note, that the bipolar coordinate system may also be seen as the stereographic projection of the points on a spherical surface onto a plane through a fixed point on the sphere's equator\cite{Papavassiliou2017}.

\subsection{Separation Ansatz}
Based on the separability of the Laplace equation [\cref{eq:expansion_laplace_eq_conforma_trafo}] and the media-wise treatment, we consider the formal solution 
 \begin{align}
 V^{(\alpha)}(s,w) 
   = \sum\limits_{n=1}^{\infty} 
        f^{(\alpha)}_n(s) ~ g^{(\alpha)}_n(w)
 \label{app-eq:V_separ_ansatz_general_coord}
 \end{align}
 where we have directly discarded the $n=0$ contribution. 
The solution separates into ''radial'' functions
\begin{align}
f^{(\alpha)}_n(s) =   A^{(\alpha)}_n \tilde{R}^n(s) 
                    + B^{(\alpha)}_n \tilde{R}^n(-s)
\label{app-eq:general_f_alpha_n}
\end{align}
with 
\begin{align}
\tilde{R}(\zeta) = \exp(-\zeta)  
    \quad \text{and} \quad
\tilde{R}(\rho)  = \exp(+\rho)    
\end{align}
and ''angular'' functions
\begin{align}
 g_n(w) = C_n \cos(nw) + D_n \sin(nw)
\label{app-eq:general_g_alpha_n}
\end{align}
The latter function is directly considered indepent of the the particular medium $\alpha$, as the angular dependence is directly determined by the external potential via the far-field limit $V_0(\mu,\nu)$ given by \cref{app-eq:V_0_log_pol,app-eq:V_0_bipo}. We find
\begin{align}
A^{(2N)}_n = A^{\rm inc}_n = -\tilde{E}_0 ~~,~~
C_n = \cos\gamma ~~ \text{and} ~~  D_n = \sin\gamma
\end{align}
where 
\begin{align}
 \tilde{E}_0 = \left\{ 
                 \begin{matrix}
                         {\rm log-polar~coordinates}: &  2 a E_0           \\
                    {\rm bipolar~coordinates}: &  \delta_{n1} E_0
                 \end{matrix} 
               \right.
 \label{app-eq:E_0_exp_coeffs}              
\end{align}
The remainder will be dedicated to the analytical determination of the coefficients $A^{(\alpha)}_n$ and $B^{(\alpha)}_n$ for the different kinds of wires.
 
\subsection{\label{app-sec:potential_core_shell} Core-single-shell nanowires}
We start with the core-single-shell wires. The treatment of nonconcentric realizations of the latter via conformal transformations has been suggested in Ref.\cite{Aubry2013}. For a fully electrostatic problem, the capacitance has been derived in Ref.~\cite{LuchtBipolar}. In Ref.~\cite{Zhang13} a solution is presented with a correction which approximately accounts for radiative losses. 
Let us also note, that the nonconcentrically coated cylinder has also been treated within Ref.\,\cite{Starkov2017}, where the effective dielectric landscape of a random, but parallel arrangement of such wires in some host material has been approximated based on Maxwell-Garnett theory, without reference to potential plasmonic resonances.

Setting the core as a metal with Drude permittivity $\epsilon_0=\epsilon(\omega)$ while the shell and environment are described by constants, we deduce from the boundary conditions [\cref{eq:continuous_norm_diel_displ,eq:continuous_tang_electric_field}] and \cref{app-eq:general_f_alpha_n} the coefficients
\begin{align}
	A^{(0)}_n & = - \tilde{E}_0 \frac{4\epsilon_1\epsilon_2}{\epsilon_1+\epsilon_2} 
	\frac{1}{ (\epsilon(\omega) - \epsilon_1)\tilde{\epsilon}_{12}(\tilde{R}_0^{2n}/\tilde{R}_1^{2n}) + \epsilon(\omega) + \epsilon_1 } 
	\label{app-eq:core_shell_A_0_n}\\
	A^{(1)}_n & = - \tilde{E}_0 \frac{2\epsilon_2}{\epsilon_1+\epsilon_2} 
	\frac{\epsilon(\omega) + \epsilon\subshell}{ (\epsilon(\omega) - \epsilon_1)\tilde{\epsilon}_{12}(\tilde{R}_0^{2n}/\tilde{R}_1^{2n}) + \epsilon(\omega) + \epsilon_1} 
	\label{app-eq:core_shell_A_1_n} \\
	B^{(1)}_n & = \tilde{E}_0 \frac{2\epsilon_2}{\epsilon_1+\epsilon_2} 
	\frac{\epsilon(\omega) - \epsilon_1}{ (\epsilon(\omega) - \epsilon_1)\tilde{\epsilon}_{12}(\tilde{R}_0^{2n}/\tilde{R}_1^{2n}) + \epsilon(\omega) + \epsilon_1 } ~R^{2n}_0 
	\label{app-eq:core_shell_B_1_n} \\
	B^{(2)}_n & = \tilde{E}_0 \tilde{R}_1^{2n} \frac{\epsilon(\omega) [\tilde{\epsilon}_{12} +(\tilde{R}_0^{2n}/\tilde{R}_1^{2n})]+ \epsilon_1 [\tilde{\epsilon}_{12} -(\tilde{R}_0^{2n}/\tilde{R}_1^{2n})]}{ (\epsilon(\omega) - \epsilon_1)\tilde{\epsilon}_{12}(\tilde{R}_0^{2n}/\tilde{R}_1^{2n}) + \epsilon(\omega) + \epsilon_1} 
	\label{app-eq:core_shell_B_e_n}
\end{align}
with $\tilde{\epsilon}_{ij} = (\epsilon_i-\epsilon_j)/(\epsilon_i+\epsilon_j)$ and the short notation $\tilde{R}_\alpha = \tilde{R}(\mu_\alpha)$.

To alleviate the derivation of the dispersion relation of dielectric-core-metal shell resonances, we provide the scattered field coefficients
for $\epsilon_2=\epsilon(\omega)$ being a Drude metal such that a convenient formulation reads 
\begin{align}
	A^{(0)}_n & = 4\epsilon_2 \tilde{E}_0 \phantom{\tilde{R}^2_0} 
	\frac{\epsilon(\omega) }{\epsilon^2(\omega) + \epsilon(\omega) (\epsilon_0 + \epsilon_2) \eta^{(0,1)}_n + \epsilon_0\epsilon_2} \\
	A^{(1)}_n & = 2\epsilon_2 \tilde{E}_0 \phantom{\tilde{R}^2_0}  \frac{\epsilon(\omega) + \epsilon_0}{\epsilon^2(\omega) + \epsilon(\omega) (\epsilon_0 + \epsilon_2) \eta^{(0,1)}_n + \epsilon_0\epsilon_2} \\
	B^{(1)}_n & = 2\epsilon_2 \tilde{E}_0 \tilde{R}_0^2 \frac{\epsilon(\omega) - \epsilon_0}{\epsilon^2(\omega) + \epsilon(\omega) (\epsilon_0 + \epsilon_2) \eta^{(0,1)}_n + \epsilon_0\epsilon_2} \\
	B^{(2)}_n & = \tilde{E}_0 \tilde{R}_1^2 
	 \frac{\epsilon(\omega) [\epsilon(\omega)- \epsilon_2 \eta^{(0,1)}_n] + }{ \epsilon^2(\omega) + \epsilon(\omega) (\epsilon_0 + \epsilon_2) \eta^{(0,1)}_n +}   \nonumber \\
	 & \quad~\quad~\quad~ \frac{+ \epsilon_0 [\epsilon(\omega) \eta^{(0,1)}_n - \epsilon_2]}{ + \eta^{(0,1)}_n + \epsilon_0\epsilon_2} 
	\label{app-eq:di_core_met_shell_B_e_n}
\end{align}
with
\begin{align}
  \eta^{(0,1)}_n 
       = \frac{1 + (\tilde{R}_0/\tilde{R}_1)^{2n}}
              {1 -  (\tilde{R}_0/\tilde{R}_1)^{2n}} 
\end{align}
We note that the assumption of local electromagnetic response in each medium allows for a straightforward correspondence between the number of boundary conditions at each interface and the number of expansion coefficients, or equivalently, independent potential contributions, within each region. Introducing nonlocal response, as is typically required for metallic regions, modifies this balance by introducing additional field components inside the corresponding medium and additional boundary conditions at the interfaces enclosing it. Consequently, a direct mapping between metal-core--dielectric-shell and dielectric-core--metal-shell configurations is no longer straightforward. A related complication arises when modified boundary conditions are employed to approximately incorporate nonlocal effects, as these approaches alter the interface conditions albeit without introducing the additional internal field contributions associated with a fully nonlocal description.


\subsection{\label{app-sec:potential_multi_ring} Core-multi-shell nanowires}
Following partly the strategy of Ref.~\cite{YarivYeh1984}, we derive the expansion coefficients of the different media applying the transfer matrix method. In comparison to the reference, a stack of multiple plane layers is found only in the transformed frame, i.e. in either log- or bipolar space. Also, we consider periodicity along the interfaces. Further, we do not consider retardation.
 
As regards the core-multi-shell wires, we consider $N$ shell units of fixed radial width $\Delta R$ (from the concentric view point) consisting of a dielectric followed by a metallic shell with differing permittivities $\epsilon_I=\epsilon_{\rm d}$ and $\epsilon_{II}=\epsilon(\omega)$.
Similar to the construction of the horizontal cross-section of the plasmonic doppler grating\cite{See2017}, we parametrize the radii of the bipolar bull's eye wires as
\begin{align}
 R_{2\xi}   & = R_0 +  \xi    \Delta R \\
 R_{2\xi-1} & = R_0 + (\xi-1) \Delta R + \beta \Delta R  
\end{align}    
where $\xi{\in}\{1, 2, \dots, N\}$ iterates the $N$ shell units. Also, we have introduced the variable thickness $0\le \beta \Delta R\le \Delta R$ of odd shells to slightly modify the geometry relative to the value of $\beta=1/2$ used in Ref.~\cite{See2017}. To fix $\Delta R$, we set the largest radius to a value that is reconcilable with the quasistatic approximation and fix the core radius as well as the amount of shell units.  With that we find  
\begin{align}
\Delta R = \frac{R_{2N} - R_0}{N}
\end{align} 
Note, that $N$ must not be too large as otherwise the macroscopic Maxwell's equations must not be applied to the media anymore. To fully specify the geometry and focal parameter, we set the center displacement $\Delta x_1$. The center coordinates then follow from \cref{app-eq:focal_param_via_one_circle,app-eq:x_c_alpha_via_radii_and_cent_displ}. The radial and center displacements of the shells are constrained by the (multi)-crescent limit 
\begin{align}
(1-\beta)\Delta R \ge \Delta x_{2\xi} 
\quad \textrm{and} \quad
\beta \Delta R \ge \Delta x_{2\xi-1}
\end{align}
To solve for the expansion coefficients, we change the notation of $(A^{(\alpha)}_n, B^{(\alpha)}_n)$ for the shells, following 
\begin{align}
     A^{(\alpha=2\xi-1)}_n
        \mapsto
           A^{(\xi,{\rm I})}_{n} 
        \quad \text{and} \quad   
     A^{(\alpha=2\xi)}_n
         \mapsto A^{(\xi,{\rm II})}_{n}     
\end{align}
and similar for the $B^{(\alpha)}_n$.
~Evaluating the boundary conditions at $s=s_0$ we find
\begin{align}
\colvec{{A}^{(1,I)}_{n}}
{{B}^{(1,I)}_{n}}
=  & 
\colvec{{A}^{(1)}_n}
{{B}^{(1)}_n} 
= \underline{\underline{T^{0\to 1}_n}} 
\colvec{{A}^{(0)}_n}
{0}
\end{align}
with
\begin{align}
\underline{\underline{T^{0\to1}_n}} 
= \frac{\epsilon_{\rm I} + \epsilon_{0}}{2\epsilon_{\rm I}} 
\mat{1 \phantom{\tilde{R}_0^{2n}}}
{\tilde{\epsilon}_{{\rm I},0} {\tilde{R}_0^{-2n}}}
{\tilde{\epsilon}_{{\rm I},0} {\tilde{R}_0^{2n}}}
{1}
\end{align}
For the transfer from the odd to even layer in a given shell unit, i.e., at 
$s = s_{2\xi-1}$ we find 
\begin{align}
\colvec{{A}^{(\xi, II)}_{n}}
{{B}^{(\xi,II)}_{n}} 
= \underline{\underline{T^{(\xi, {\rm I\to II})}_{n}}} 
\colvec{{A}^{(\xi,I)}_{n}}
{{B}^{(\xi,I)}_{n}}
\end{align}
where 
\begin{align}
\underline{\underline{T^{(\xi,{\rm I\to II})}_{n}}} 
& = \frac{\epsilon_{II} + \epsilon_{I}}
{2\epsilon_{II}}
\mat{                          1   }
{ \tilde{R}_{2\xi-1}^{-2n} \tilde{\epsilon}_{II,I} }
{ \tilde{R}_{2\xi-1}^{2n}  \tilde{\epsilon}_{II,I} }
{                              1 }                                      
\end{align}
At the transition between shell units, i.e., at 
$s = s_{2\xi}$ we find
\begin{align}
\colvec{{A}^{(\xi+1,I)}_{n}}
{{B}^{(\xi+1,I)}_{n}} 
= \underline{\underline{T^{(\xi, {\rm II\to I})}_{n}}} 
\colvec{{A}^{(\xi,II)}_{n}}
{{B}^{(\xi,II)}_{n}}
\end{align}
where 
\begin{align}
\underline{\underline{T^{(\xi, {\rm II\to I})}_{n}}} 
& = \frac{{\epsilon}_{II} + {\epsilon}_{I}}
{2\epsilon_{I}}
\mat{ \phantom{\tilde{R}_{2\xi}^{-2n}} 1  }
{          \tilde{R}_{2\xi}^{-2n}  \tilde{\epsilon}_{I,II} }
{          \tilde{R}_{2\xi}^{2n}   \tilde{\epsilon}_{I,II} }
{ \phantom{\tilde{R}_{2\xi}^{-2n}} 1  }
\end{align}
This defines the shell unit translation matrix
\begin{align}
\underline{\underline{T^{(\xi)}_n}}  
  = \underline{\underline{T^{(\xi, {\rm II\to I})}_{n}}} ~
    \underline{\underline{T^{(\xi, {\rm I\to II})}_{n}}} 
\label{app-eq:unit_cell_translation_matrix}
\end{align}
connecting the odd media for which we find
\begin{widetext}
	\begin{align}
	\underline{\underline{T^{(\xi)}_n}}
	= \frac{1}{4\epsilon\subdiel\epsilon(\omega)}
	\mat{
		g^{(2\xi-1,2\xi)}_{+,n}(\omega)	
	}
	{
		c_+(\omega)c_-(\omega) 
		\mu^{(2\xi-1,2\xi)}_{-,n}
		\tilde{R}^{-2n}_{2\xi-1}
	}
	{
		-c_+(\omega)c_-(\omega) 
		\mu^{(2\xi-1,2\xi)}_{-,n}
		\tilde{R}^{2n}_{2\xi}	
	}
	{
		g^{(2\xi-1,2\xi)}_{-,n}(\omega)
		(\tilde{R}^{2n}_{2\xi}
		/ \tilde{R}^{2n}_{2\xi-1})
	}
	\end{align}
\end{widetext}
The matrix determinant equals unity as expected from a matrix connecting two media of same dielectric constant. The remaining matrix describes the transition between the scatterer and environment, given by boundary conditions at 
$s=s_{2N}$ according to
\begin{align}
\colvec{{A}^{(2N+1)}_{n}}
{{B}^{(2N+1)}_{n}} =
\underline{\underline{T^{2N\to 2N+1}_n}}
\colvec{{A}^{(N,II)}_{n}}
{{B}^{(N,II)}_{n}}
\end{align}         
where
\begin{align}
\underline{\underline{T^{2N\to 2N+1}_n}} 
&=
\frac{{\epsilon}_{2N+1} + {\epsilon}_{II}}
{2{\epsilon}_{2N+1}}
\mat{ 1 }
{ \tilde{R}_{2N}^{-2n} \tilde{\epsilon}_{2N+1,II} }
{ \tilde{R}_{2N}^{2n}  \tilde{\epsilon}_{2N+1,II} }
{ 1 }
\end{align}
As such, we can connect the coefficients of the core and environment following  
\begin{align}
\colvec{{A}^{\rm inc}_{n}}
{{B}^{(2N+1)}_{n}} =
\underline{\underline{G_n}}
\colvec{{A}^{(0)}_{n}}
{0}
\end{align}
which defines the overall transfer matrix $\underline{\underline{G_n}}$ as
\begin{align}
\underline{\underline{G_n}} =
\underline{\underline{T^{2N\to 2N+1}_n}} ~~
\underline{\underline{T^{(N, {\rm I\to II})}_{n}}} ~~
\left[\prod\limits_{\xi=1}^{N-1}\underline{\underline{T^{(\xi)}_n}}\right] ~~
\underline{\underline{T^{0\to 1}_n}}
\end{align}
Note, that the product here corresponds to matrix multiplication from the left so that $\underline{\underline{T^{(1)}_n}}$ appears to the right.
As such, we obtain the scattered field coefficient
\begin{align}
B^{(2N+1)}_n = ( G_{n,21} / G_{n,11}) A^{\rm inc}_n 
\label{app-eq:multi_ring_B_e_n}
\end{align}
from which all other coefficients can be obtained by means of a proper choice of individual transfer matrices. Notice, that the latter are similar in form, differing only in the actual media and radii appearing in the respective indices given the purely local description without modified boundary conditions.
\section{Stability of the cylinder-core resonance under shell addition}

In \cref{sec:multi_ring_systems} we have observed the well-known quasistatic cylindrical surface plasmons [$\epsilon(\omega){=}-\epsilon\subdiel$] independent of the amount of shell units (including $N{=}0$) provided a metallic cylindrical wire builds the core and the dielectrica are of same material given by $\epsilon\subdiel$. 
Here we prove this observation based on the transfer matrix method. For simpliticity we set here also $\epsilon_{2N+1}{=}\epsilon\subdiel$, to rewrite the total transfer matrix as  
\begin{align}
\underline{\underline{G_n}} = 
\left( \prod\limits_{\xi=2}^N \underline{\underline{T^{(\xi)}_n}} 
\right)~
\underline{\underline{T^{(1)}_n}}~\underline{\underline{T^{0\to 1}_n}} 
\label{app-eq:redef_transfer_mat_for_cyl_res_proof}
\end{align}   
where we have isolated the ''tube-wire cavity'' from the structure whose denominator we have already factorized into a  cylindrical and modified ring modes [\cref{eq:G_mat_coeff_frac_tube_wire}]. Note further, that the product here corresponds to matrix multiplication from the left so that $\underline{\underline{T^{(2)}_{n}}}$ appears to the right.

Commuting the limit with the finite product in \cref{app-eq:redef_transfer_mat_for_cyl_res_proof}, we consider
eventually 
\begin{align}
\prod\limits_{\xi=2}^N \lim\limits_{\epsilon(\omega)\to-\epsilon\subdiel}\underline{\underline{T^{(\xi)}_n}} &= 
\mat{\chi^{2n}}
{0}
{0}
{\chi^{-2n}} \\
\text{with} \quad & \chi = \prod\limits_{\xi=2}^N \left(\tilde{R}_{2\xi-1} / 
\tilde{R}_{2\xi}\right) 
\label{app-eq:limit_of_redef_transfer_mat_at_cyl_res_proof}
\end{align}   
As such we only need the first column of the tube-wire matrix
given by the entries
\begin{align}
\left(T^{(1)}T^{0\to 1}_n\right)_{11} 
 &= 
 c_+ \left[ g_{n,+}^{(1,2)} 
            - (c_-)^2 \mu_{n,-}^{(1,2)} (\tilde{R}^{2n}_0 
             /\tilde{R}^{2n}_1 ) \right]
 \\
\left(T^{(1)}_nT^{0\to 1}_n\right)_{21}
&= 
 -c_- \left[ (c_+)^2\mu_{n,-}^{(1,2)} 
            +  g_{n,-}^{(1,2)} (\tilde{R}^{2n}_0/\tilde{R}^{2n}_1) \right] \tilde{R}^{2n}_2
\end{align} 
so that finally we have
\begin{align}
B^{(2N+1)}_n \approx  
\frac{\tilde{E}_0(\tilde{R}^{2n}_2 / \tilde{R}^{2n}_1) ~\chi^{-4n} \tilde{R}^{2n}_0}
{(\tilde{R}_1^{2n} / \tilde{R}^{2n}_2) 
	+ (\tilde{R}_0^{2n} / \tilde{R}_1^{2n} - \tilde{R}_0^{2n} / \tilde{R}^{2n}_2)} \frac{c_+}{c_-}
\end{align}
to leading order in the limit $\epsilon(\omega)\to-\epsilon\subdiel$, which holds the 
cylindrical pole independent of the summaiton index of the field expansion. 

\section{\label{app-sec:abs_eff}Derivation of absorption efficiency}
Within the manuscript, we have considered the absorption efficiency 
as the observable of choice. Here, we derive it analytically, 
detailing the important steps for the bipolar coordinates.
For completeness, we eventually present the analogous results for the 
log-polar coordinates. 

The absorption efficiency $Q_{\rm abs}$ derives from the
rate at which electromagnetic energy crosses a control volume containing the scatterer\cite{bohren2008absorption}. 
Due to the vertical symmetry, the (infinite) height $H$ of the wires eventually cancels such that we consider directly the rate $P_{\rm abs}$ per unit length given a control circle as boundary.
This rate is given by
\begin{align}
P_{\rm abs}(\omega) &= -\int_{\Ccont} \vec{dC} \cdot \Re\left[ \vec{E} \times \left(\vec{H}\right)^* \right]
\label{app-eq:P_abs_noncon}
\end{align} 
representing the surface-normal projection of the Poynting vector, which is assumed to be a monochromatic field contribution of frequency $\omega$ and we average over the period
$P = 2\pi / \omega$. 

For the nonconcentric structures, we choose a control circle with $\zeta < \zeta_{2N}$. In the following, we restrict to the right halfspace ($\zeta>0$). Therein the outward pointing surface element reads $H\vec{d\Ccont}= Hh(u,\zeta)[-\hat{\zeta}(u,\zeta)]du$.  The fields are decomposed into the external (subscript $0$) and scattered field (superscript $s$).
The external field is embodied by a linearly polarized plane wave
\begin{align}
\vec{E_0}(\vec{r}) & = E_0 \left( \cos\gamma \hat{x} + \sin\gamma \hat{y}\right) \exp(i \vec{k}\cdot \vec{r}) \\ 
\vec{H_0}(\vec{r}) & = \sqrt{\frac{\epsilon_0}{\mu_0}} E_0 \left( \hat{k} \times \hat{E_0} \right)               \exp(i \vec{k}\cdot \vec{r}) 
\end{align}  
given also in-plane propagation $\vec{k} = k_x \hat{x} + k_y \hat{y}$. 
Taking the the quasistatic limit yields
\begin{align}
\vec{E_0}(\vec{r}) & = E_0 \left( \cos(\tau) \hat{x} + \sin(\tau) \hat{y}\right)
\label{app-eq:uniform_E_field}\\ 
\vec{H_0}(\vec{r}) & = \sqrt{\frac{\epsilon_0}{\mu_0}} E_0 \left( \hat{k} \times \hat{E_0} \right) = H^0_z \hat{z}    
\label{app-eq:uniform_H_field}  
\end{align}
The bare contribution due to the external fields vanishes due to the spatial homogeneity
of their cross product. This is
in line with the nature of free space propagation. The Poynting vector aligns with the 
wave vector. So, on a technical level, the above statement boils down to a calculation of the 
integrals 
\begin{align}
\mathcal{I}_{0,x/y} & = \int_0^{2\pi} du ~ h(\zeta,u) \zeta_{x/y}(\zeta,u)
\end{align}
The integral $\mathcal{I}_{0,x}$ can be shown to vanish for $\zeta>0$ by residue theorem mapping it onto a contour integral along the unit circle which includes the two-fold pole $z=\exp(-\zeta)$. 
Further,  $\mathcal{I}_{0,y}$ vanishes due to the odd parity of the integrand and its $2\pi$-periodicity. 

From the potentials derived in \cref{app-sec:potential_ring_structures}, we can infer the scattered fields based on their conservative nature [\cref{eq:static_faraday_law}]. Therein, we leave the outermost expansion coefficients $B^{(2N+1)}_n$ (where we substitute $2N+1\to2$ in the following for a single shell) unspecified to present all considered structures equally. For $\zeta>0$ the components of the scattered electric field can be brought into the form
\begin{align}
E^{\rm s}_{z, n} (\zeta,u)  & = 0 
\label{app-eq:bipo_E_scatt_z} \\
E^{\rm s}_{\zeta, n} (\zeta,u) 
&=   -n E_0 B^{(2N+1)}_n h^{-1}(\zeta,u) \exp(n\zeta) \times \nonumber \\
 & \quad \times \left[ \cos(\tau) \cos(nu) + \sin(\tau) \sin(nu) \right] 
\label{app-eq:bipo_E_scatt_rho}\\
E^{\rm s}_{u, n} (\zeta,u) 
&=   +n E_0 B^{(2N+1)}_n h^{-1}(\zeta,u) \exp(n\zeta) \times \nonumber \\
   & \quad \times \left[ \cos(\tau) \sin(nu) - \sin(\tau) \cos(nu) \right]     
\label{app-eq:bipo_E_scatt_u}         
\end{align}

Note the finite multipolar contributions brought about by the nonconcentric arrangement of 
rings inherent in the finiteness of the $B^{(2N+1)}_n$ for $n>1$.
The scattered magnetic field is obtained from the Amp\`ere-Maxwell equation, which
expanded in bipolar coordinates yields 
\begin{align}
 \partial_\zeta {H}^{\rm s}_z 
        & =          -  i \omega \epsilon_0 \epsilon_{2N+1} h(\zeta,u) {E}^{\rm s}_u \\
 \partial_u    {H}^{\rm s}_z 
        & = \phantom{-} i \omega \epsilon_0 \epsilon_{2N+1} h(\zeta,u) {E}^{\rm s}_\zeta \\
 \partial_\zeta [h(\zeta,u) H^{\rm s}_u] 
        &=  \phantom{-} \partial_u [h(\zeta,u) H^{\rm s}_\rho]   
\end{align}
We neglect the horizontal components and consider a Fourier series expansion
of the $z$-component to find 
\begin{align}
 H^{\rm s}_{z,n}
     &= (-i\omega \epsilon_0\epsilon_{2N+1})
            E_0 B^{(2N+1)}_n \exp(n\zeta) \times \nonumber \\
     & \quad \times      
            \left[ \cos(\tau) \sin(nu) - \sin(\tau) \cos(nu) \right]
\label{app-eq:bipo_H_scatt_z}            
\end{align}
From \cref{app-eq:bipo_E_scatt_rho,app-eq:bipo_E_scatt_u,app-eq:bipo_E_scatt_z,app-eq:bipo_H_scatt_z}
it can be deduced, that 
\begin{align}
 & \hat{\zeta} \cdot \Re \left\{        \vec{E}^{\rm s}(\vec{r}, \omega) 			
 \times 
 \left[\vec{H}^{\rm s}(\vec{r}, \omega)\right]^{*} 
 \right\} \propto \Re \left\{ i\right\} = 0
\end{align}
signifying the vanishing radiation loss that we expect from a leading-order
quasistatic calculation\cite{Ford1984}. 
  
For the remaining contributions, we first consider the conversion rate tied to the scattered 
electric field scaling with the integral
\begin{align}
\int^{2\pi}_0 du ~ h(\zeta,u)~ \hat{\zeta} \cdot & \left[ \vec{E}^{\rm s} \times \left(\vec{H_0}\right)^{*}\right] 
 \nonumber \\
& = \left(H_{0,z}\right)^* \int^{2\pi}_0 du ~ h(\zeta,u) E^{\rm s}_u    
\end{align}
where we have used the field definitions and the cross products of the unit vectors. By \cref{app-eq:bipo_E_scatt_u}, the metric factor cancels and the resulting integrals of trigonometrics for each order $n$ vanish.
As such, even a finite $\vec{H_0}$ does not contribute to leading order quasistatic energy conversion. This is also true for the log-polar coordinates, where the integral evaluation is more straightforward due to the simpler metric factors.
We are left with
\begin{align}
P_{\rm abs} = \int^{2\pi}_0 du ~ h(u,\zeta) ~\hat{\zeta} \cdot \Re \left\{ \vec{E_0} \times \left(\vec{H}\supsct \right)^*\right\}   
\end{align}
Essentially, the solution depends on the evalutation of the integrals 
\begin{align}
  \mathcal{I}^{~\mathcal{T} }_{\alpha,n} 
   &= 
      \int^{2\pi}_0 du ~ h(u,\zeta) \zeta_\alpha(u,\zeta) \mathcal{T}(nu)
\end{align}
due to the definition of fields and unit vectors wherein $\mathcal{T}\in\{\cos;\sin\}$ .
By means of parity and periodicity of the integrands, it can be deduced that
$\mathcal{I}_{x,n}^{\sin}=\mathcal{I}_{y,n}^{\cos}=0$. To solve the remaining integrals, we map to an integration over the unit circle and apply the residue theorem with a two-fold pole $z=\exp(-\zeta)$ to find
\begin{align}
  \mathcal{I}^{\cos}_{x,n}
        &  = 
  \mathcal{I}^{\sin}_{y,n} 
         =          -  2n  \exp(-n\zeta)   
\end{align}
As such we find 
\begin{align}
P_{\rm abs} &= 2a \omega \epsilon_0 \epsilon_{2N+1} |E_0^2| \sum\limits_{n=1}^{\infty} \Im \left\{ n B^{(2N+1)}_n\right\} 
\end{align}
which after normalization by the geometrical cross section (per unit length) $\sigma_{\rm geom}/H{=}2R_{2N}$ and external field intensity $I_0{=} \sqrt{\epsilon_0\epsilon_{2N+1}/\mu_0}|E_0|^2/2$ yields the absorption efficiency
\begin{align}
  Q_{\rm abs} &= \frac{2\pi k_{2N+1} a}{R_{2N}}\sum\limits_{n=1}^{\infty} \Im \left\{ n B^{(2N+1)}_n\right\}
\label{app-eq:Q_abs_noncon_via_B_e_n}  
\end{align}
with environmental wave number $k_{2N+1} = \sqrt{\epsilon_{2N+1}}\omega/c_0$.
Note, that the variable $\zeta$, defining the control circle, does not enter the result, since the surrounding
medium is not absorbing. Further, the result does not depend on the angle of field polarization (This has been observed also for tube-wire cavities \cite{YuLuo_2012}). However, we clearly observe how the nonconcentricness affects the amount of excited multipoles. More technically, the independence 
of external field polarization is a consequence of residence of all nonconcentric boundaries in a single 
halfspace. As a counter example, consider the treatment of the cylindrical dimer\cite{Moeferdt2018,nonloc_dimer_paper} which necessitates both sets of $\zeta-$coordinate lines and is tied to polarization-dependent
selection rules for the plasmonic modes. We note however, that the charge and field distributions generally depend on the polarization direction.

For the concentric structure, we find
\begin{align}
Q_{\rm abs} &= \frac{\pi k_{2N+1}}{R_{2N}}\Im \left\{ B^{(2N+1)}_1\right\}
\label{app-eq:Q_abs_con_via_B_e_n}
\end{align}
where now only dipolar modes are excited. This provides an extension of the bare cylinder efficiency presented, e.g., in Ref.\,\cite{Crotti_2022}.

Let us emphasize, again, that the above contour techniques assume that $\zeta > 0$ as otherwise, 
the poles of the integrands can lie on the circle $|z|=1$ where $z=\exp(iu)$, for which we cannot apply the residue theorem. This particular case, which is necessary for the crescent, need to be discussed 
separately. See, for instance, Refs.~\cite{Aubry_2010_NanoLett,Aubry_2010_NanoLett,Aubry_2010_PRB} using a different coordinate transform. Other than in our consideration, there does not appear a periodic variable.

Lastly, let us derive some valuable limits at the example of the core-single-shell wire which we interchange with that of the modal sum in \cref{app-eq:Q_abs_noncon_via_B_e_n}. First of all, we reproduce the concentric limit for which $\Delta x_1=0$, given fixed $R_0$ and $R_1$. According to \cref{app-eq:x_c_alpha_via_radii_and_cent_displ} 
\begin{align}
 \lim\limits_{\Delta x_1\to 0} x_0 \to \infty
\end{align}
We note, that the decisive terms due to the definition of $B^{2}_n$ following \cref{app-eq:core_shell_B_e_n} behave as
\begin{align}
    \lim\limits_{\Delta x_1\to0} a^2 \exp(-2n \zeta_{0/1}) \sim x^2_0 \left(\frac{R_{0/1}}{2x_{{0/1}}} \right)^{2n} 
\label{app-eq:concentric_limit_Q_abs_series_term_scaling}
\end{align} 
where we have used
\begin{align}
    & \lim\limits_{\Delta x_1\to0} \exp(-\zeta_{0/1}) 
     = \lim\limits_{x_{0}\to\infty} \exp[-\arccosh(x_{0/1}/R_{0/1})]  \nonumber \\
    & \sim \exp[-\ln(2x_{0/1}/R_{0/1})] = \frac{R_{0/1}}{2x_{0/1}}
\label{app-eq:eff_rad_a_Dx_to_zero}    
\end{align}
As such, in \cref{app-eq:concentric_limit_Q_abs_series_term_scaling}, only the term with $n=1$ does not vanish. This is in line with the mere excitation of the dipolar mode in the leading-order quasistatic approximation. With that we reproduce the absorption efficiency for a concentric system as in \cref{app-eq:Q_abs_con_via_B_e_n}. 

In the opposite limit of maximal $\Delta x_1 = R_1-R_0$, we recreate a crescent shape. We then have 
\begin{align}
\lim\limits_{\Delta x_1\to R_1 - R_0} x_{0/1}   =  R_{0/1}   
\end{align}
With that we find 
\begin{align}
\lim\limits_{\Delta x_1\to R_1- R_0} \exp(-2n\zeta_{0/1})   =  1                 
\end{align}
As such, the expansion coefficients are basically scaling with the focal parameter, so that
\begin{align}
\lim\limits_{\Delta x_1\to R_1 - R_0} B^{(1)}_n 
\propto
\lim\limits_{\Delta x_1\to R_1 - R_0} a = 0
\label{app-eq:B_e_n_limit_a_to_zero}
\end{align}
Accordingly, the localized surface plasmons we have derived here, do not contribute in this limit.

\section{\label{sec:Mie_theory}Mie theory of the concentric bull's eye wire}
\begin{figure}[h]
	\centering
	\includegraphics[scale=0.33]{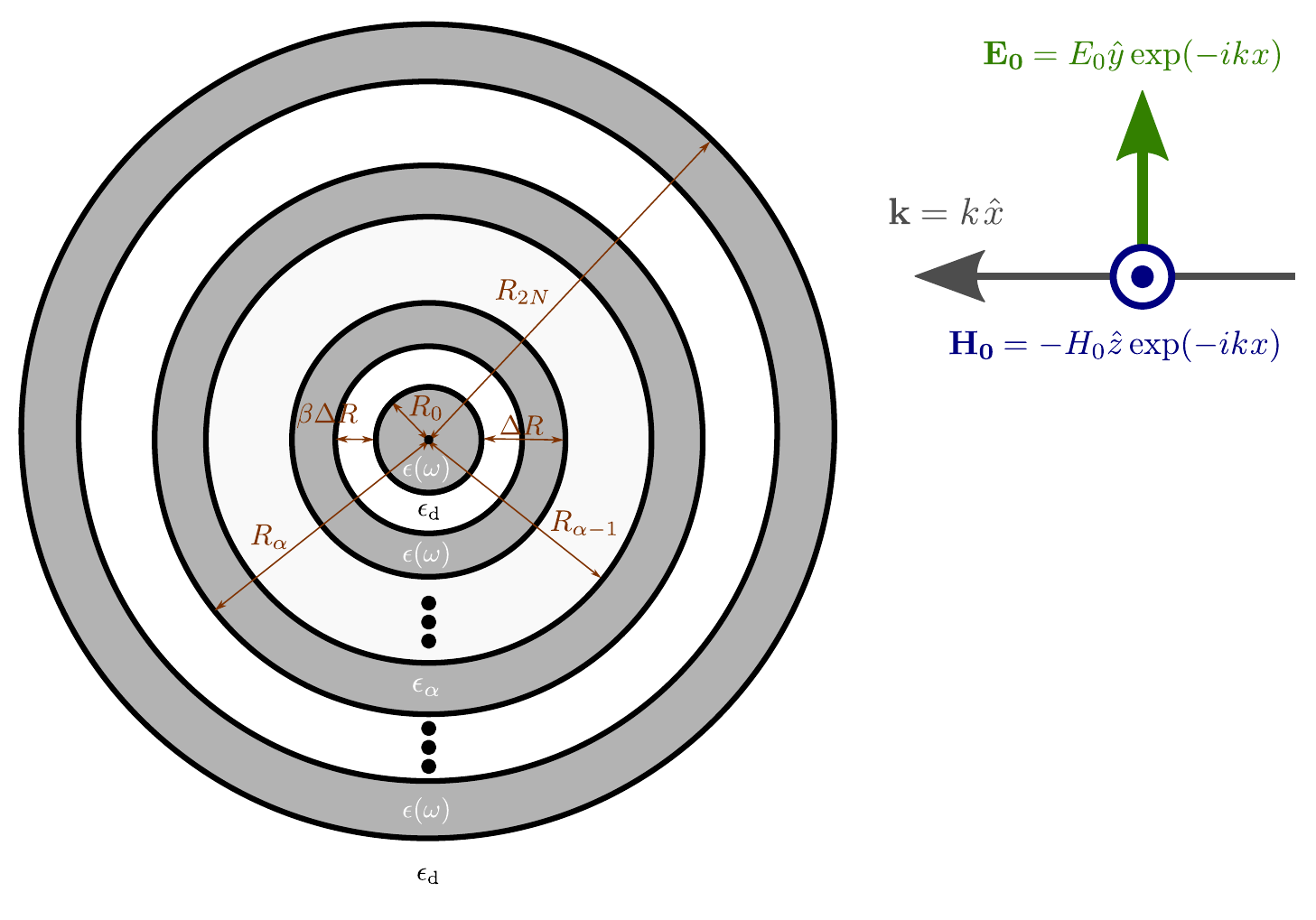}
	\caption{
		\label{fig:Mie_sketch_con_bulls_eye}
		Cross-sectional sketch of an infinitely extended, concentric bull's eye wire excited by a plane wave propagating along the negative $x$ direction and polarized perpendicular to the rotational axis ($\hat{z}$). Owing to the translational invariance along the wire axis, the electromagnetic problem reduces to a two-dimensional description. 
	}
\end{figure}

Within \cref{subsec:multi_ring_compare_doppler} we employ the Mie theory in order to obtain the fully retarded solution of the scattering problem which further accounts for the propagating portions of the fields. Here we aim to present a Mie theory of a concentric bull's eye setup for arbitrary number $N$ of shell units, where $N=1$ corresponds to the tube-wire cavity (see \cref{fig:Mie_sketch_con_bulls_eye}). 

We follow the derivations of Ref.~\cite{bohren2008absorption} for a single cylinder under axis-normal incidence and linear-polarization of a plane wave, now amended by a sequence of $N$ shell units of alternating dieletric and Drude-metal shells. For oblique incidence on a multilayered cylinder, see for instance Ref.\,\cite{Li2000}, where the focus lies on chiral materials.

The fields are expanded in cylindrical coordinates using the well-known vector cylindrical harmonics \cite{bohren2008absorption}
\begin{align}
\vec{M}_n = \nabla \times \left( \hat{z} \psi_n \right)
\quad  \text{and} \quad 
\vec{N}_n = (1/k) \nabla \times \vec{M}_n
\end{align}
where $k$ is the wavenumber in the respective medium, in which $\psi_n$ solves the scalar Helmholtz equation and 
where $n$ is the mode order given by azimuthal periodicity of the potentials and thus fields.
The scalar solution builds on the basis functions 
\begin{align}
\psi_n(r, \theta, z) &= \mathcal{B}_n(k\sqrt{r^2-h^2}) \exp(in\theta) \exp(ihz)
\end{align}
where $r$, $\theta$ and $z$ are the cylindrical radius, azimuthal angle and wire axis coordinate while $\mathcal{B}_n$ denotes the choice of cylindrical Bessel function. 

Given the transverse response throughout and for a $y$-polarized incident plane wave propagating along the negative $x$-direction, which is orthogonal to the wire axis along $z$, the field of layer $\alpha$ constrained by $R_{\alpha-1}{<}r{<}R_\alpha$ (where $R_{-1}{=}0$) may conveniently be written
\begin{align}
\vec{E}^{(\alpha)}
&= \frac{iE_0}{k_\alpha} \sum_{n=-\infty}^{\infty} \left(-i\right)^n  
\left[  a^{(\alpha)}_n \vec{M}^{\mathcal{J}}_n(k_{\alpha} r) 
+ b^{(\alpha)}_{n} \vec{M}^{\mathcal{H}}_n(k_{\alpha} r) \right] \label{app-eq:Mie_bulls_eye_general_E_expansion}\\
\vec{H}^{(\alpha)}
&= \frac{E_0}{\omega} \sum_{n=-\infty}^{\infty} \left(-i\right)^n 
\left[    a^{(\alpha)}_{n} \vec{N}^{\mathcal{J}}_n(k_{\alpha} r)   
+ b^{(\alpha)}_{n} \vec{N}^{\mathcal{H}}_n(k_{\alpha} r) 
\right] 
\label{app-eq:Mie_bulls_eye_general_H_expansion}
\end{align} 
where $\mathcal{J}$ and $\mathcal{H}$ denote the cylindrical Bessel and Hankel function of first kind, respectively.
For finiteness, inside the core we have $b^{(0)}_n{=}0$. It can be shown, that the incident field corresponds to the coefficients $a^{(2N+1)}_n{=}-1$ 
and for normal incidence $h{=}0$, which is inherited to the other fields by the boundary conditions.

\begin{figure*}
	\centering
	\includegraphics[scale=.4]{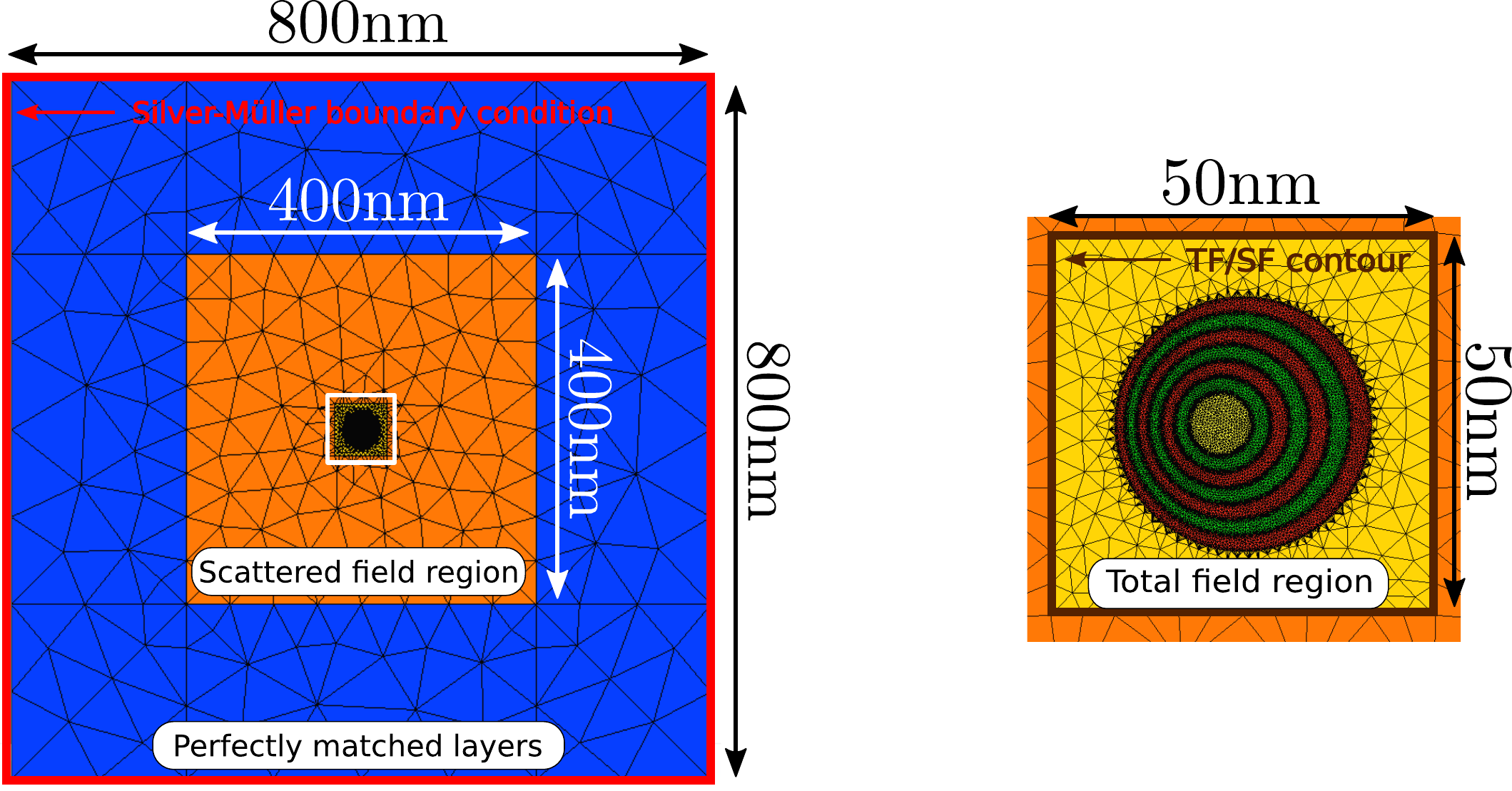}
	\caption{
		\label{fig:finest_mesh_bulls_eye_noncon}
		Depiction of the finest mesh employed for the simulations of 
		the bipolar bull's eye wire, highlighting the PML region (blue), 
		scattered field region (orange) as well as the total field region 
		(yellow; inside white square boundary with zoom on the right) 
		containing the scatterer (yellow core ; green and red shells). The 
		total field scattered field contour (brown line) and outer boundary 
		(red) for Silver-Müller boundary condition are also highlighted.}	
\end{figure*}

As with the quasistatic problem the vectors of field expansion coefficients in each layer are two-dimensional. They are fixed by the typical Maxwell boundary conditions, wherein we neglect infinitesimally thin current sheets. Among these we are only interested in the continuity of the surface tangential components of the electric and magnetic field. For normal incidence, we are interested into the continuity of $E_\theta$ and $H_z$. These are determined by \cref{app-eq:Mie_bulls_eye_general_E_expansion,app-eq:Mie_bulls_eye_general_H_expansion} and 
\begin{align}
\vec{M}_n^\mathcal{B} 
&= \left[ - k \mathcal{B}'(kr) \exp(i n \theta)\right]    
\hat{\theta} \\
\vec{N}_n^\mathcal{B} 
&= \left[ k \mathcal{B}(kr) \exp(i n \theta)\right]    
\hat{z}
\end{align}
where the prime denotes the derivative with respect to the argument 
and we find for two neighboring media $\alpha$ and $\alpha+1$
\begin{align}
\colvec{ a^{(\alpha+1)}_n}
{ b^{(\alpha+1)}_n}
&=
\underline{\underline{T^{(\alpha)}_n}}
\colvec{ a^{(\alpha)}_n}
{ b^{(\alpha)}_n}       
\end{align}
with 
\begin{widetext}
	\begin{align}
	\left(T^{(\alpha)}_n\right)_{00}
	&= 
	\frac{   k_{\alpha\phantom{+1}} \mathcal{J}_n(k_{\alpha\phantom{+1}}R_\alpha)\mathcal{H}'_n(k_{\alpha+1}R_\alpha) 
		- k_{\alpha+1} \mathcal{H}_n(k_{\alpha+1}R_\alpha)\mathcal{J}'_n(k_{\alpha\phantom{+1}}R_\alpha) }
	{   k_{\alpha+1} \mathcal{J}_n(k_{\alpha+1}R_\alpha)\mathcal{H}'_n(k_{\alpha+1}R_\alpha) 
		- k_{\alpha+1} \mathcal{H}_n(k_{\alpha+1}R_\alpha)\mathcal{J}'_n(k_{\alpha+1}R_\alpha)
	} \\
	\left(T^{(\alpha)}_n\right)_{01}
	&= 
	\frac{   k_{\alpha\phantom{+1}} \mathcal{H}_n(k_{\alpha\phantom{+1}}R_\alpha)\mathcal{H}'_n(k_{\alpha+1}R_\alpha) 
		- k_{\alpha+1} \mathcal{H}_n(k_{\alpha+1}R_\alpha)\mathcal{H}'_n(k_{\alpha\phantom{+1}}R_\alpha) }
	{   k_{\alpha+1} \mathcal{J}_n(k_{\alpha+1}R_\alpha)\mathcal{H}'_n(k_{\alpha+1}R_\alpha) 
		- k_{\alpha+1} \mathcal{H}_n(k_{\alpha+1}R_\alpha)\mathcal{J}'_n(k_{\alpha+1}R_\alpha)
	}	    	\\
	\left(T^{(\alpha)}_n\right)_{10}
	&=      
	\frac{   k_{\alpha\phantom{+1}} \mathcal{J}_n(k_{\alpha\phantom{+1}}R_\alpha)\mathcal{J}'_n(k_{\alpha+1}R_\alpha) 
		- k_{\alpha+1} \mathcal{J}_n(k_{\alpha+1}R_\alpha)\mathcal{J}'_n(k_{\alpha\phantom{+1}}R_\alpha) }
	{   k_{\alpha+1} \mathcal{H}_n(k_{\alpha+1}R_\alpha)\mathcal{J}'_n(k_{\alpha+1}R_\alpha) 
		- k_{\alpha+1} \mathcal{J}_n(k_{\alpha+1}R_\alpha)\mathcal{H}'_n(k_{\alpha+1}R_\alpha)
	}	   \\      
	\left(T^{(\alpha)}_n\right)_{11}
	&=
	\frac{   k_{\alpha\phantom{+1}} \mathcal{H}_n(k_{\alpha\phantom{+1}}R_\alpha)\mathcal{J}'_n(k_{\alpha+1}R_\alpha) 
		- k_{\alpha+1} \mathcal{J}_n(k_{\alpha+1}R_\alpha)\mathcal{H}'_n(k_{\alpha\phantom{+1}}R_\alpha) }
	{   k_{\alpha+1} \mathcal{H}_n(k_{\alpha+1}R_\alpha)\mathcal{J}'_n(k_{\alpha+1}R_\alpha) 
		- k_{\alpha+1} \mathcal{J}_n(k_{\alpha+1}R_\alpha)\mathcal{H}'_n(k_{\alpha+1}R_\alpha)
	}	   
	\end{align}
\end{widetext}
and where $a^{(2\xi-1)}_n=a^{(\xi,I)}_n$ and $a^{(2\xi)}_n=a^{(\xi,II)}_n$ with the associated radii and materials.
Via iteration, we can express the coefficients of the environment by those of the core to find
\begin{align}
\colvec{a_n^{(2N+1)}}
{b^{(2N+1)}_n}
&= 
\underline{\underline{G_{n}}}
\colvec{a_n^{(0)}}
{0} 
~~\text{with}~~
\underline{\underline{G_{n}}}
= \prod\limits_{\alpha=0}^{2N} \underline{\underline{T^{(\alpha)}_{n}}}           
\end{align}
where the product symbol corresponds to matrix multiplication from the left so $\underline{\underline{T^{(0)}_{n}}}$ appears on the right.
The derived scattered field coefficient can then be introduced into the definitions of the fully retarded 
scattering  and extinction efficiencies reading
\begin{align}
Q\subsct    &= \frac{2}{k_{2N+1} R_{2N}} \sum\limits_{n=-\infty}^{\infty}                        |b^{(2N+1)}_n|^2 \\
Q_{\rm ext} &= \frac{2}{k_{2N+1} R_{2N}} \sum\limits_{n=-\infty}^{\infty} \operatorname{Re}\left\{b^{(2N+1)}_n\right\}
\end{align}
yielding the absorption efficiency $Q_{\rm abs}= Q_{\rm ext} - Q\subsct$. Employing corresponding individual transfer matrices, the fields can also be calculated.

Lastly, we note, that for a nonlocal response within the shell of a core-shell structure the Mie theory is presented in Ref.\,\cite{raza213_nanotube}. 
Further, multilayer spheres have been treated within Mie theory. A short history is provided in \cite{zhang2025}.

\begin{table*}
	\centering
	\begin{ruledtabular}	
		\begin{tabular}{l|cccccc}
			geometrical cross section & CPU & cores & max RAM & smallest in-circle radius & time step & simulation time \\
			bipolar bull's eye (N=3)
			& AMD EPYC 7282 16-Core
			& 8
			& 125GB
			&  0.04nm
			&  0.25as
			& 19d 19h 15min 6s \\
			Doppler bull's eye (N=3)
			& AMD EPYC 7313 16-Core Processor
			& 8
			& 251Gb
			& 0.05nm
			& 0.26as 
			& 15d 5h 41min 27s                  
		\end{tabular}
		\caption{\label{tab:simulation_information} Summary of the computational specificities necessary for an estimation of the cost of the simulations for the finest simulations of the nonconcentric bull's eye wires.}	
	\end{ruledtabular}
\end{table*}	

\section{\label{app-sec:num_method} Numerical simulations}
As another means to assess the optical response of the different wires considered within the manuscript, we employ the Discontinuous Galerkin time domain (DGTD) method to solve Maxwell's equations numerically. The spatial discretization is based on a finite-element mesh. We create finite-element meshes representing a space-filling, conformal and unstructured assembly of triangles, which can be locally adapted with respect to shape and size for sufficient approximation of curved interfaces. For this we employ the open-source software GMSH\cite{gmsh_article}. Element-local function spaces allow for an efficient calculation of field expansions as they allow for considerable parallelization of element-local linear operations, excluding, however, operations regarding inter-element communication. We employ a nodal method based on Ref.~\cite{gmsh_article} for direct field value extraction. The DGTD method represents a time-domain method so that we can naturally inject a light pulse of finite temporal width and extract different spectral contributions by Fourier methods applied to the outcome of a single simulation run. Employing a total-field/scattered-field (TF/SF) method, we inject a plane wave with Gaussian envelope  defined by a carrier (angular) frequency $7.44\,$eV (about $0.8$ times the plasma frequency of silver) and temporal FWHM $1.57\,$fs propagating along the positive $x$-axis and being linearly polarized along the $y$-axis. The spectral information is obtained by an on-the-Fly Fourier transform. All simulations have covered a physical duration of $667.128\,$fs each to ensure a sufficient decay of the scattered field. Otherwise, an abrupt termination of the simulated fields will introduce spurious oscillations into the Fourier-transformed quantities (Gibbs phenomenon).   
The DGTD allows for a large class of time steppers. We rely on the low-storage explicit Runge Kutta method of 14 stages and fourth order\cite{niegemann_efficient_2012} which provides a convenient trade-off between accuracy and stable time step.
Lastly, the focus of the DGTD method on balance equations allows for a natural introduction of the Drude model by which we describe the conduction electronic response within the metal. 
We focus on silver based on the refractive index data of Ref.~\cite{Johnson1972}. The plasma frequency amounts to $\omega_{\rm p}=9.149\,$eV with a Drude damping rate of $\gamma=0.021\,$eV. The background permittivity is set to unity.
To mimic an open system we apply a perfectly matched layer (PML) around the physical domain which is terminated by a quadratic contour on which we apply a Silver-Müller boundary condition.
For an application oriented introduction of the DGTD method, see  \cite{lpor_DGTD_review}. An exemplary mesh of the bipolar bull's eye (N=3) wire considered in \cref{fig:bulls_eye_vs_doppler_spectra} is presented in \cref{fig:finest_mesh_bulls_eye_noncon}. The smallest element features an in-circle radius of $0.04\,$nm. \Cref{tab:simulation_information} summarizes the necessary information to estimate the requirements of the spatially most discretized simulations of the nonconcentric bull's eye wires employed in \cref{fig:bulls_eye_vs_doppler_spectra}.

~\newpage
\bibliographystyle{apsrev4-2}
\bibliography{nonconcentric}

\end{document}